\documentclass[11pt,preprint]{aastex}

\begin{document}

\shorttitle{HD163296: it's a gas}
\shortauthors{Rosenfeld et al.}

\title{A Spatially Resolved Vertical Temperature Gradient in the HD 163296 Disk}

\author{
Katherine A. Rosenfeld\altaffilmark{1},
Sean M. Andrews\altaffilmark{1}, \\
A. Meredith Hughes\altaffilmark{2},
David J. Wilner\altaffilmark{1},
\& Chunhua Qi\altaffilmark{1}
}
\altaffiltext{1}{Harvard-Smithsonian Center for Astrophysics, 60 Garden Street, Cambridge, MA 02138}
\altaffiltext{2}{Van Vleck Observatory, Astronomy Department, Wesleyan University, 96 Foss Hill Drive, Middletown, CT 06459, USA}

\begin{abstract}
We analyze sensitive, sub-arcsecond resolution ALMA Science Verification 
observations of CO emission lines in the protoplanetary disk hosted by the 
young, isolated Ae star HD 163296.  The observed spatial morphology of the 
$^{12}$CO $J$=3$-$2 emission line is asymmetric across the major axis of the 
disk; the $^{12}$CO $J$=2$-$1 line features a much less pronounced, but similar,
asymmetry.  The $J$=2$-$1 emission from $^{12}$CO and its main isotopologues 
have no resolved spatial asymmetry.  We associate this behavior as the direct 
signature of a vertical temperature gradient and layered molecular structure in 
the disk.  This is demonstrated using both toy models and more sophisticated 
calculations assuming non-local thermodynamic equilibrium (non LTE) conditions. 
A model disk structure is developed to reproduce both the distinctive spatial 
morphology of the $^{12}$CO $J$=3$-$2 line as well as the $J$=2$-$1 emission 
from the CO isotopologues assuming relative abundances consistent with the 
interstellar medium.  This model disk structure has $\tau=1$ emitting surfaces 
for the $^{12}$CO emission lines that make an angle of $\sim 15^\circ$ with 
respect to the disk midplane.  Furthermore, we show that the spatial and 
spectral 
sensitivity of these data can distinguish between models that have 
sub-Keplerian gas velocities due to the vertical extent of the disk and its 
associated radial pressure gradient (a fractional difference in the bulk gas 
velocity field of $\gtrsim 5$\%).
\end{abstract}

\section{Introduction}

Spectral line emission from molecules can be a powerful tool for studying the 
structure of protoplanetary disks.  Abundant molecules in disks, such as CO or 
CN, trace the spatial distribution of gas temperatures and the radial extent of 
the molecular hydrogen that composes most of the disk mass 
\citep{koerner93,guilloteau13}.  Analysis of multiple lines can determine the 
vertical temperature structure of the disk \citep{dartois03} or indicate 
freezeout of the gas phase molecules onto dust grains in the cold midplane 
\citep{qi11}.  Observations of various species are useful for studying the 
abundance distributions of molecules and processes like ionization 
\citep{oberg11}, grain surface reactions \citep{dutrey11}, and fractionation 
\citep{oberg12}.  Furthermore, these spectral lines probe the bulk motions, or 
kinematics, of the gas \citep{beckwith93}.  Observations from interferometers 
are particularly useful since the emission is resolved both spatially and 
spectrally and can be used to derive a dynamical mass of the young star 
\citep{guilloteau98,simon00}, probe the structure of the disk 
\citep{dutrey08,panic10}, or detect non-thermal line broadening from turbulence 
\citep{hughes11,guilloteau12}.

The Herbig Ae star HD 163296 (spectral type A1) hosts a protoplanetary disk with
bright continuum emission \citep{allen76,mannings97,natta04} and a rich
molecular spectrum \citep{qi01,thi01,thi04,qi13}.  It is an isolated system, not
known to be associated with any star forming region or young moving cluster 
\citep{finkenzeller84}.  Observations of CO lines show that the disk exhibits
Keplerian rotation \citep{mannings97,isella07,hughes08} and is remarkably large,
extending past a radius of $\sim 500$\,AU \citep{isella07}.  The disk is seen in
scattered light \citep{grady00} and is associated with an asymmetric Herbig-Haro
outflow \citep{devine00} and molecular disk wind \citep{klaassen13}.  The system
is young \citep[$\sim 5$\,Myr;][]{natta04} and its {\it Hipparcos} parallax 
indicates it is nearby \citep[$d= 122^{+17}_{-13}$\,pc;][]{vandenancker98}. 
\citet{qi11} developed a density and temperature structure for this disk that 
was consistent with Submillimeter Array (SMA) observations of both the dust 
continuum and multiple CO emission lines.  Including an abundance distribution 
that accounts for both freezeout and photodissocation, their analysis suggested 
that the disk has a midplane CO snow line at $r\sim155$\,AU.  An ancillary 
result of their model is that the disk is colder in the midplane than in the 
atmosphere where the higher $J$ emission lines are produced.  Evidence for a 
vertical temperature gradient in this disk was independently presented by 
\citet{akiyama11}.

We present an analysis of the spectral line emission from the main CO 
isotopologues observed by the Atacama Large Millimeter/sub-millimeter Array 
(ALMA) towards the HD 163296 disk in Bands 6 and 7.  We first use a toy model 
comprised of a rotating double cone to guide our intuition of this data. 
Leveraging the high spatial and spectral resolution of these observations, we 
then demonstrate that the vertical temperature gradient in the disk is directly 
resolved in the $^{12}$CO $J$=3$-$2 maps alone.  We develop a disk structure 
model that is roughly self-consistent with those data and the $J$=2$-$1 
transitions of $^{12}$CO, $^{13}$CO, and C$^{18}$O.  In addition, we explore the
velocity structure of the disk and show that the excellent spatial and spectral 
resolution of the sensitive Band 7 observations can distinguish between various 
classes of disk models which have bulk gas velocities that deviate at the 
$\gtrsim 5$\% level.  We describe the observations in \S\ref{sec:observations} 
and show the results in \S\ref{sec:results}.  Our analysis follows in 
\S\ref{sec:analysis}, where we present the vertical temperature gradient 
(\S\ref{sec:models}) and explore sub-Keplerian motions due to the vertical 
geometry of the disk and its radial pressure gradient (\S\ref{sec:velocity}).  
We conclude with a brief discussion in \S\ref{sec:discussion} and a summary in 
\S\ref{sec:summary}.

\section{Observations and Data Reduction}\label{sec:observations}

HD 163296 ($\alpha = 17^{\rm h} 56^{\rm m} 21\fs287$, 
$\delta = -21\degr57\arcmin22\farcs39$, J2000) was observed by ALMA in Band 6 
and Band 7 as part of its commissioning and science verification (SV) program. 
The raw and calibrated visibility data were publicly released through the 
science portal, and were accompanied by a set of detailed calibration and 
imaging scripts provided by the ALMA SV team\footnote{\url{https://almascience.nrao.edu/alma-data/science-verification/overview}}.  
We started with the calibrated measurement set and used the CASA software 
package \citep[v3.4;][]{mcmullin07} to produce self-calibrated and continuum 
subtracted spectral visibilities for the $^{12}$CO, $^{13}$CO, C$^{18}$O 
$J$=2$-$1 lines at 230.538, 220.399, 219.560\,GHz in Band 6 and $^{12}$CO 
$J$=3$-$2 at 345.796\,GHz in Band 7.

The Band 6 observations were taken using 24 of the ALMA 12\,m antennas on 2012 
June 9, June 23, and July 7 for a total on-source time of 84 minutes (including 
latency).  The baselines spanned a range of 20 to 400\,m.  The correlator was 
configured to simultaneously observe four spectral windows (SpW), with two in 
each sideband:  SpWID \#1 included the $^{13}$CO and C$^{18}$O $J$=2$-$1
transitions, while SpWID\#2 covered the  $^{12}$CO $J$=2$-$1 line.  SpWIDs \#0 
and \#3 observed line-free continuum \citep[but see][]{qi13} and were centered 
at 217 and 234\,GHz.  The flux calibrator for each of the nights was Juno, 
Neptune, and Mars, respectively, and the bandpass solution was generated from 
observations of the quasar J1924-292.  Integrations on the science target were 
interleaved with $\sim 2$ minute long observations of the nearby quasar 
J1733-130 for phase calibration.  For the July 7 observations, the flux density 
of J1733-130 was determined to be 1.55\,Jy, using a bootstrap from the flux 
calibrator, Mars.  Similarly, HD 163296 was observed in Band 7 on 2012 June 9, 
Jun 11, June 22, and July 6 with the same antenna configuration.  Juno (June 9 
only) and Neptune were observed as primary flux calibrators.  When Neptune was 
observed, the flux density of J1733-130 was bootstrapped for each individual 
SpW.  An additional bootstrap from these observations was used to determine the 
flux density of J1733-130 in each SpW observed on June 9th.  The total 
on-source time was 140 minutes.  These observations used the same bandpass and 
phase calibration strategy as the Band 6 observations.  The correlator was set 
up to observe four spectral windows, with two in each sideband: SpWIDs \#0 and 
\#3 were centered at 360 and 347\,GHz for continuum, HCO$^+$ $J$=4$-$3 at 
356.734\,GHz was detected in SpWID \#1, and the $^{12}$CO $J$=3$-$2 line was 
covered by SpWID \#2.\footnote{The transformation to the absolute sky frequency 
in the Band 7 measurement set provided by the SV team is incorrect.  The 
$^{12}$CO $J$=3$-$2 line suggests that the systemic velocity, $v_{\rm sys}$, of 
HD 163296 as measured in the local standard of rest frame (radio definition; 
LSRK) is $6.99$\,km s$^{-1}$.  This disagrees with other measurements 
\citep[$v_{\rm sys}=5.8$\,km s$^{-1}$;][]{mannings97,isella07,hughes08} as well 
as the ALMA Band 6 observations.  This velocity offset was confirmed by the ALMA
Helpdesk, but is only a minor inconvenience since the line morphology is, for 
this small offset, sensitive only to the velocity differences.}

\begin{figure}[t!]
\epsscale{0.95}
\plotone{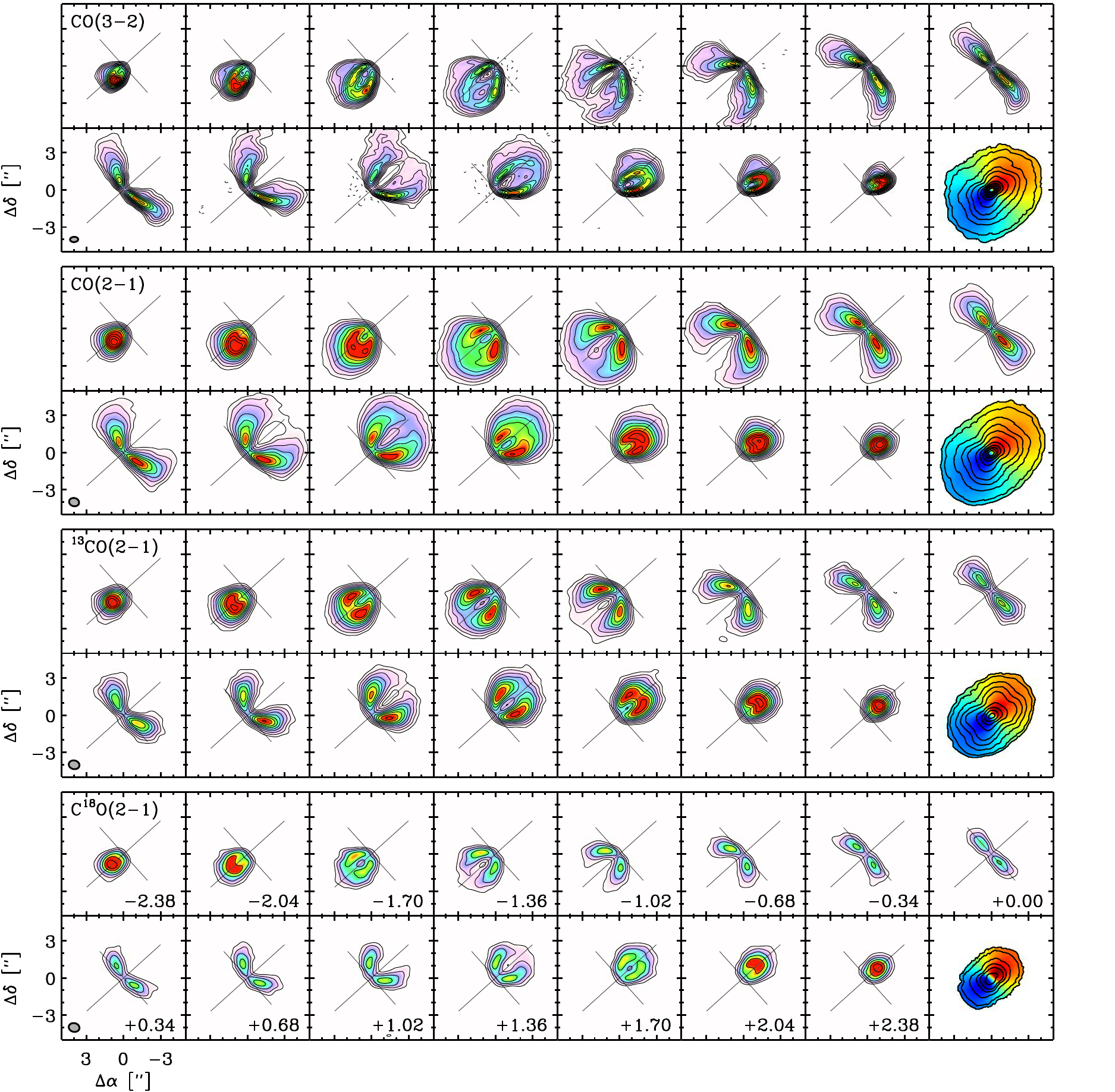}
\figcaption{The disk hosted by HD 163296 imaged in four CO lines. The 
disk orientation is indicated by the solid gray lines, the synthesized beam 
dimensions are shown in the lower left corner panels.  Each line has been 
regridded onto the same velocity resolution (written relative to the systemic 
velocity in km s$^{-1}$).  The emission is shown in both color and with 
$\sigma \times [3,6,10 + 5n]\,(n = 0,1,2,\ldots)$ contours (see Table 
\ref{tab:lines} for noise estimates).  The last panel for each line contains the
0th (contours) and 1st (color scale) integrated moment maps.
\label{fig:datachmaps}}
\end{figure}

\begin{figure}[t!]
\epsscale{0.95}
\plotone{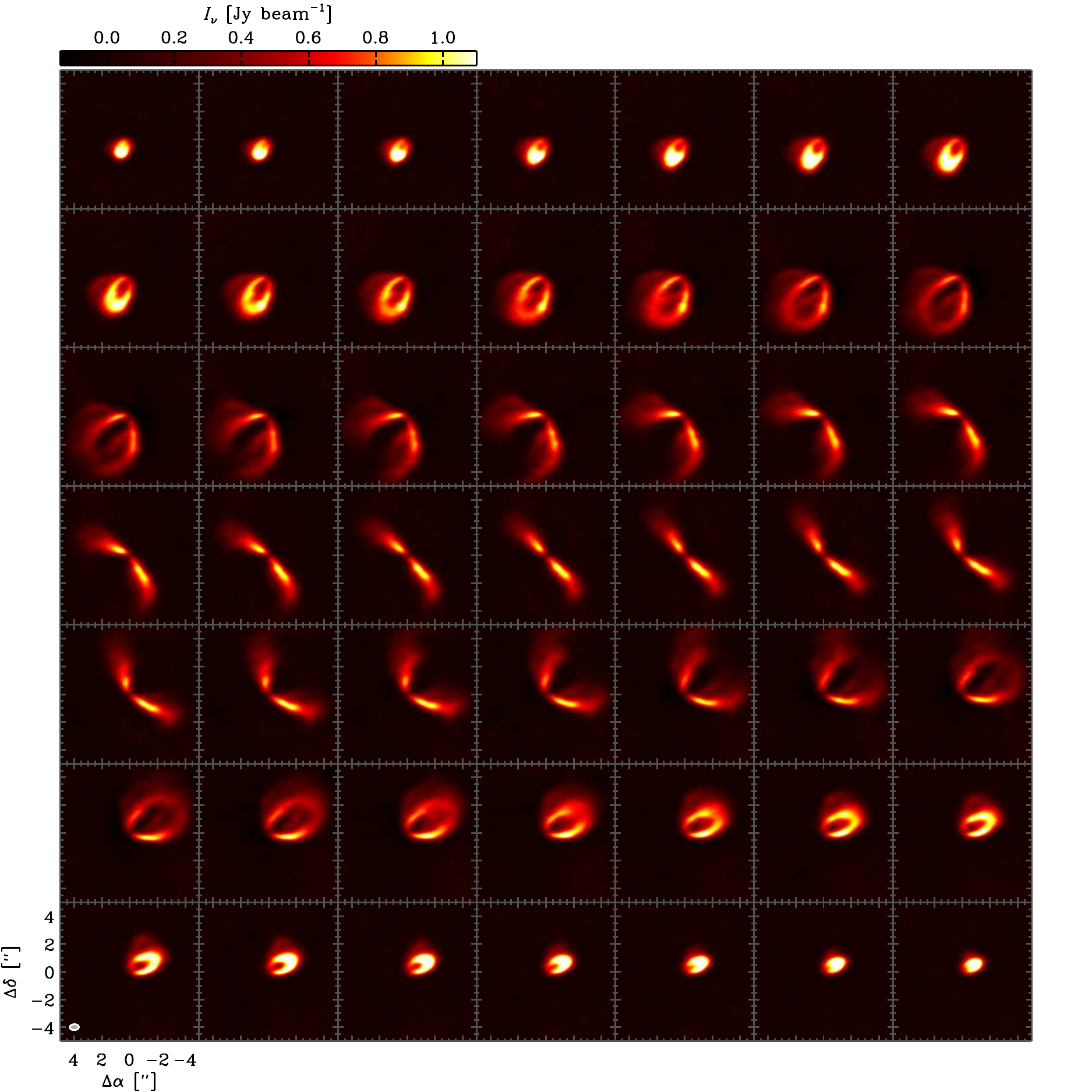}
\figcaption{The $^{12}$CO $J$=3$-$2 line imaged with a channel width of
$\Delta v = 0.11$\,km s$^{-1}$.  
\label{fig:CO32chmaps}}
\end{figure}

\section{Results}\label{sec:results}

Figure \ref{fig:datachmaps} shows channel maps of the $^{12}$CO $J$=3$-$2 line 
along with the $J$=2$-$1 lines from $^{12}$CO, $^{13}$CO, and C$^{18}$O.  
Before imaging, the visibilities have been regridded onto the same 
0.34\,km s$^{-1}$ wide channels; only the central 15 channels are shown, where 
the emission pattern is clearly resolved for all lines.  The central 49 channels
of the $^{12}$CO $J$=3$-$2 line imaged at the native 
$\Delta v=0.11$\,km s$^{-1}$ channel spacing is also shown in Figure 
\ref{fig:CO32chmaps}.  We generated these synthesized images using the CASA 
implementation of the CLEAN algorithm \citep{hogbom74} with natural weighting.  
Table \ref{tab:lines} summarizes the noise characteristics, synthesized beams, 
native channel widths and integrated intensities for each line.  The reported 
line intensities agree with previous measurements by the SMA \citep{qi11} 
within the 10\% systematic uncertainty estimated for the absolute calibration of
the visibility amplitudes.

In order to produce spectral images with such high fidelity, we developed a 
CLEAN mask based on the emission pattern expected from a disk in Keplerian 
rotation \citep{beckwith93}, which requires some assumption about the stellar 
mass, disk inclination, disk size, phase offset, position angle, and systemic 
velocity (see \S\ref{sec:toy}).  Except for the stellar mass, all of these
parameters can be estimated from an initial imaging of the data with a square 
mask; the final imaging results are consistent for any reasonable choices of 
these parameters.  We identified the region of the rotating disk where the 
projected line of sight velocities corresponded to each individual spectral 
channel.  This region was then convolved with a fixed Gaussian kernel 
($\sigma = 0\farcs7$) to broaden the mask and ensure that all of the significant
emission was covered and that the CLEANed area was always much larger than the 
synthesized beam size.  This recipe produces a separate mask for each channel, 
within which the measurement set was CLEANed deeply.  Some additional CLEAN 
iterations were then applied with no mask.  This masking technique significantly
improves the quality of the images compared to the reference maps provided in 
the SV data package (the signal-to-noise ratio per channel improves by a factor 
of $\sim$2--3).

The $^{12}$CO $J$=3$-$2 emission observed toward the disk around HD 163296 is 
asymmetric in its spatial morphology across the major axis of the disk 
(PA$=312^\circ$).  For channels offset from the line center by velocities 
$\gtrsim 1$\,km s$^{-1}$, the resolved emission appears systematically closer 
to the southern semi-minor axis (see Figure \ref{fig:CO32chmaps} and the top set
of channel maps in Figure \ref{fig:datachmaps}).  Furthermore, 
channel-by-channel, the southern half of the disk is brighter than the northern.
The $^{12}$CO $J$=2$-$1 line has these same features (refer to Figure 
\ref{fig:datachmaps}), but the effect is much less obvious; the morphological 
asymmetry can only be clearly seen for a few channels, 
$|\Delta v| \approx 1.7$--2.0\,km s$^{-1}$.  The $^{13}$CO and C$^{18}$O 
spectral emission have no apparent spatial asymmetries (see Figure 
\ref{fig:datachmaps}).

\section{Analysis}\label{sec:analysis}

Before attempting to explain the detailed morphologies of these CO data, we 
provide a brief primer on molecular line emission from a rotating disk
\citep[e.g.][]{omodaka92,beckwith93}.  The observed line is centered around some
transition frequency, $f_0$, with a linewidth that is sensitive to the 
quadrature sum of both the thermal velocity and turbulent motions of the gas.  
If the emitting gas has some bulk velocity along the line of sight, 
$v_{\rm los}$, then the line center will be Doppler shifted away from the 
transition frequency, $f_0$, by an amount $\Delta f = -v_{\rm los}(f_0/c)$.  
Observations taken by (sub-)millimeter interferometers like ALMA have both 
spectral and spatial resolution, and produce an image of the source for a set of
frequency, or equivalently velocity, channels.  These spectral line observations
comprise a three dimensional data cube of position, position, and velocity.  For
an inclined disk in Keplerian rotation, the emission observed in a single 
channel ``highlights'' what part of the disk has that same projected velocity as
it is Doppler shifted into the channel center.  

The morphology of the line emission in any given channel changes depending on 
both the physical conditions and bulk motions of the gas.  Modeling the line 
emission requires a calculation of the molecular excitation state of the gas and
an integration of the radiative transfer equation along each line of sight, $s$,
through the disk structure,
\begin{equation}
I_\nu = \int_0^\infty S_\nu(s) \exp\left[-\tau_\nu(s)\right] K_\nu(s) ds,
\label{eq:radiativetransfer}
\end{equation}
where $\tau_\nu(s) = \int_0^s K_\nu(s')ds'$ is the optical depth, $K_\nu$ is the
absorption coefficient, and $S_\nu$ is the source function.  All three of these 
terms depend upon the local temperature, $T_{\rm gas}$, and density, 
$\rho_{\rm gas}$, of the disk.  Determining the structure of a disk using 
optically thick CO lines is difficult, and so previous studies have leveraged 
observations of multiple lines to break the model degeneracies 
\citep{dartois03,pietu07}.

To parse what these data reveal about the disk hosted by HD 163296,
we will first present a toy model that qualitatively explains the observed 
morphological asymmetries noted in the $^{12}$CO transitions (\S\ref{sec:toy}).
Next, we describe a more complex modeling procedure for testing whether a given
disk structure produces the observed line intensity in addition to the 
morphological and brightness asymmetries (\S\ref{sec:models}).  We then present 
a simple model developed for pedaogogical purposes (\S\ref{sec:pedagogic}) 
followed by a more realistic hydrostatic disk model (\S\ref{sec:hydro}).  
Finally, we demonstrate that some care must be taken when defining the rotation 
pattern of the disk, in light of the disk structure we derive with its large 
radial extent and molecular layer extending relatively high above the midplane 
(\S\ref{sec:velocity}).

\subsection{A Toy Model}\label{sec:toy}

The asymmetric shape of the $^{12}$CO $J$=3$-$2 line emission cannot be 
explained by the emission pattern predicted for a vertically thin rotating disk.
The coordinates of an inclined thin disk ($x,y,z=0$), are related to the 
sky-plane relative to the observer, ($x',y'$), by
\begin{equation}
\left(\begin{array}{c} x \\ y \\ z \end{array}\right) = \left(\begin{array}{c} x' \\ y' / \cos i \\ 0 \end{array}\right)
\end{equation}
where $i$ is the disk inclination (0$^\circ$ is seen face-on).  For a disk 
rotating in circular, Keplerian motion, the projected line of sight velocity as a
function of position is
\begin{equation}
v_{\rm obs}(x',y') = \sqrt{\frac{GM_\ast}{r}} \sin i \cos \theta,
\end{equation}
with an argument $\theta = \arctan(y/x)$ and modulus $r=\sqrt{x^2+y^2}$.  This 
function is mirror symmetric across the $x'$-axis and so the shape of the 
emission observed at some velocity, $v_{\rm obs}$, should appear symmetric about
the $x'$-axis of the sky-plane.  Of course, this geometrical argument is grossly
simplified and considers neither the physical structure of the disk, nor the 
associated radiative transfer effects for spectral line emission (see 
\S\ref{sec:models}).  However, protoplanetary disks are thought to be 
geometrically thin \citep{pringle81}, with radial surface densities and 
temperatures that roughly follow power-laws \citep[i.e. slowly changing $\rho$ 
and $T$ in the outer disk;][]{weidenschilling77a,adams88}.  Under these 
conditions, this approximation works well and has been both a useful and 
effective guide in the study of line emission from protoplanetary disks.  
However, the observation analyzed here reveals an emission pattern that requires
a modified conceptual framework \citep[see also][]{pavlyuchenkov07,semenov08,
guilloteau12}.

\begin{figure}[t!]
\epsscale{0.95}
\plotone{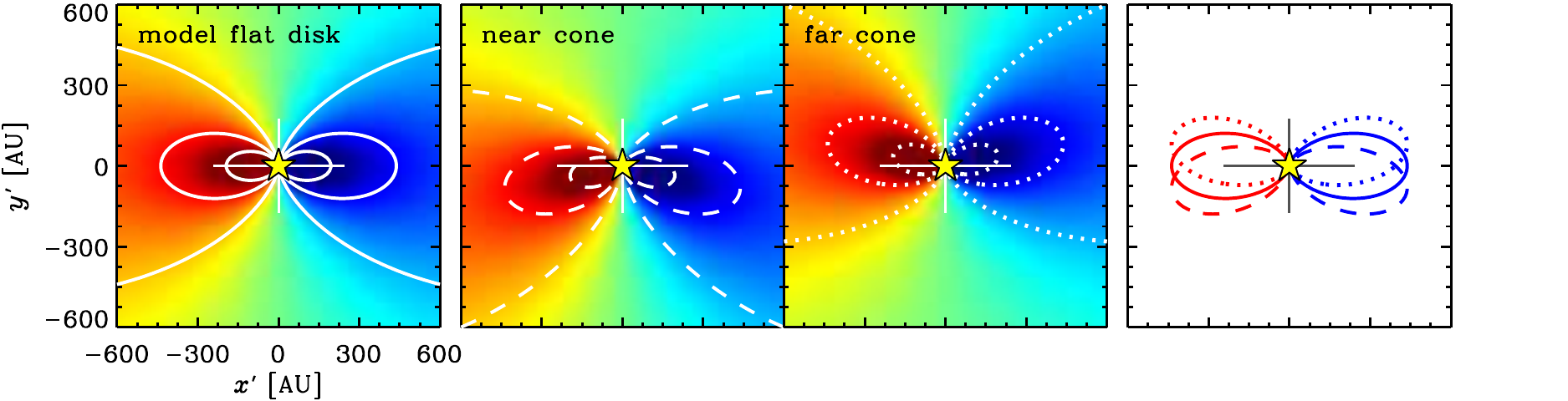}
\figcaption{The middle two panels separately show the observed velocities ({\it 
in color}) from the near (second panel) and far (third panel) halves of a double
cone structure in circular, Keplerian rotation with 
$\psi = 15^\circ$, $i=44^\circ$, 
$M_\ast = 2.3$\,M$_\odot$, and semi-major axis aligned with the $x'$-axis of the
observer (see text for parameter definitions).  Dashed contours ({\it in white})
show the $\pm$0.75, 1.5, and 2.25\,km s$^{-1}$ isovelocity contours.  The same 
quantities are shown for a flat disk ($\psi=0$; first panel).  The 
$\pm 1.5$\,km s$^{-1}$ contours for all three models are shown together in the 
fourth panel ({\it solid curve} is the flat disk, {\it dashed curve} is the near
cone, {\it dotted curve} is the far cone).
\label{fig:cone}}
\end{figure}

The toy model that we propose is a differentially rotating double cone, oriented
along the $z$ axis with an opening angle $\psi \in (0,\pi/2)$ measured from the 
$xy$ plane (i.e., the disk midplane). In this case, rays originating from the 
the sky-plane will intersect the surface at
\begin{equation}
\left(\begin{array}{c} x \\ y \\ z \end{array} \right) = \left(\begin{array}{c} x' \\\ y' / \cos i + t \sin i \\ t\cos i\end{array}\right),
\end{equation}
where $t$ solves the quadratic equation\footnote{\url{http://www.geometrictools.com/Documentation/IntersectionLineCone.pdf}}
\begin{equation}
0 = t^2 \left[\cos(2i) + \cos(2\psi)\right] - 2 \sin^2(\psi)\left[{x'}^2 + {y'}^2 \sec^2(i) + 2 t y' \tan(i)\right].
\end{equation}
The positive and negative roots of this equation correspond to the near and 
far halves of the double cone respectively (where we define the near half as 
the one closer to the observer). Any given line of sight will intersect the near
cone at {\it larger} $y$ than for the flat disk, by an additional factor 
$t\sin i$ (assuming $i < 180^\circ$).  Therefore, if this near cone is in 
circular, Keplerian rotation, ${v_\theta}^2 =GM/\sqrt{x^2 + y^2}$ with no $v_z$ 
component, the observed isovelocity contours will be systematically shifted to 
{\it smaller} $y$.   The same argument follows for the far cone, except that 
isovelocity contours are systematically shifted to {\it larger} $y$.  If 
$i>180^\circ$ (which is the case for the HD 163296 disk; see 
\S\ref{sec:models}), then the effects are the same except the solutions switch 
to describe the opposite half of the double cone.  Larger values of $\psi$ will 
amplify the differences between these curves.  The emission morphology predicted
from this double cone structure clearly mimics the observed asymmetry in the 
$^{12}$CO $J$=3$-$2 line, as demonstrated in Figure \ref{fig:cone}.

\begin{figure}[t!]
\epsscale{0.95}
\plotone{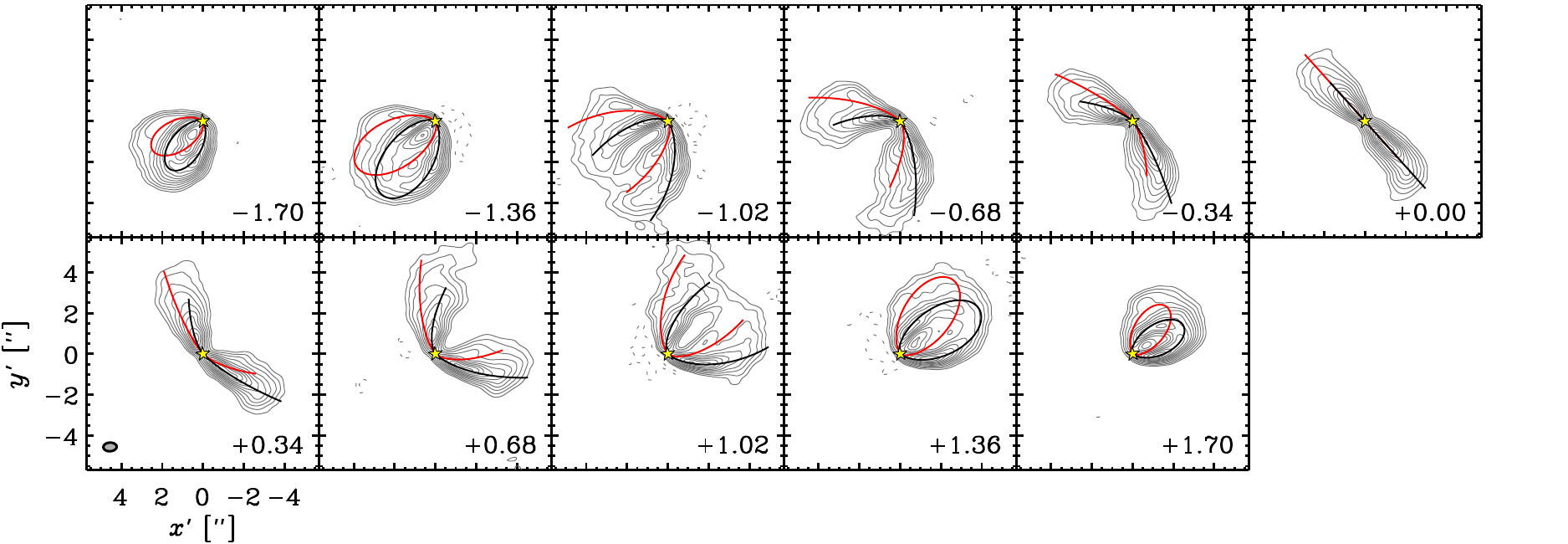}
\figcaption{The solid curves show the isovelocity contours of the toy model,
corresponding to the channel velocities of the $^{12}$CO $J$=3$-$2 line 
({\it shown in gray}).  Both the near ({\it in black}) and far ({\it in red})
halves of the double cone structure are shown out to a radius of 600\,AU.  The 
model parameters are the same as in Figure \ref{fig:cone}.
\label{fig:cone_chmaps}}
\end{figure}

The brightness and morphological asymmetries of the $^{12}$CO $J$=3$-$2 and 
$^{12}$CO $J$=2$-$1 lines can be understood if the $\tau=1$ emitting surface is 
something like a double cone structure with $\psi \sim 15^\circ$.  The bright 
portion of the line corresponds to the near cone that is systematically shifted 
along the minor axis (to smaller $y$).  The dim ``shadows'', which are 
asymmetric in the opposite sense, represent the far cone.  Figure 
\ref{fig:cone_chmaps} shows the velocity pattern of this toy model overlaid on 
the regridded channel maps of the $^{12}$CO $J$=3$-$2 emission.   The key 
feature of the double-cone is that it mimics an emission structure that has 
little contribution from the disk midplane: the structure is in some 
observational sense ``hollow'' at its center.  Furthermore, the angle that the
cone makes with the disk midplane ($\psi \sim 15^\circ$) is fairly large 
\citep[but not abnormally so;][]{vanzadelhoff01} and suggests that the disk 
cannot be treated as geometrically thin.

Detailed radiative transfer and chemical models 
\citep[][]{aikawa99,aikawa06,semenov11,walsh12} predict a cone-like structure 
for molecules like CO.  In these calculations, the midplane temperatures are 
coupled to the cold dust, greatly reducing the CO gas-phase abundance via 
freezeout onto dust grains.  Higher in the disk, direct stellar radiation heats 
the material, liberating the CO from the grain surfaces and ensuring that 
CO is both abundant and warm.  Above this molecular layer, where gas column 
densities are low, the CO is depleted by photodissociation via energetic photons
from the star and background radiation field.  It is this first layer, the cold 
midplane, that naturally explains the ALMA observations.  In this region, the 
gas is too cold to densely populate the higher $J$ CO transitions, so the 
observed emission is low.  Additionally, the CO number densities should be 
heavily depleted by freezeout onto dust grains, further reducing the emission.  
Instead, most of the emission is from the warm molecular layer which, being high
above the midplane, should produce asymmetric emission with respect to the major
axis of the disk.  This scenario also explains why the less abundant $^{13}$CO 
and C$^{18}$O emission is not asymmetric: these lines originate from deeper 
layers in the disk, which correspond to lower $\psi$ and less severe asymmetries
that are not resolved by the observations.  Indeed, calculations by 
\citet{pavlyuchenkov07} and \citet{semenov08} of HCO$^+$ $J$=4$-$3 emission for 
axisymmetric, inclined disks ($i\sim60^\circ$) predicted this same morphology.

\subsection{Physical Model Analysis}\label{sec:models}

While the toy model of a differentially rotating double-cone provides a useful
qualitative explanation for the asymmetric morphology of the $^{12}$CO $J$=3$-$2
line emission, we now undertake a more physically motivated, quantitative 
analysis.  We build upon the disk structures developed by \citet{dartois03} and 
\citet{qi11} to reproduce the morphology of the $^{12}$CO $J$=3$-$2 line along 
with the line intensities of the $J$=2$-$1 emission from the $^{12}$CO, 
$^{13}$CO, and C$^{18}$O isotopes.  In this section, we introduce the modeling 
scheme that we will use to illustrate the sensitivity of the observations to the
temperature structure and velocity field of the disk.

Following the formalism explained by \citet{andrews12}, axisymmetric density 
and temperature structures are defined using a polar cylindrical 
coordinate system ($r,z$).  The gas temperature structure, $T_{\rm gas}(r,z)$,
is specified parametrically throughout this work.  The gas surface density 
profile is assumed to be the self-similar model of a thin, viscous accretion 
disk 
\citep{lynden-bell74,hartmann98},
\begin{equation}
\Sigma_{\rm gas}(r) = \Sigma_c \left(\frac{r}{r_c}\right)^{-\gamma} \exp \left[-\left(\frac{r}{r_c}\right)^{2-\gamma}\right],
\end{equation}
where $r_c$ sets the radial size of the gas disk, $\gamma$ is a power-law index,
and $\Sigma_c = M_{\rm gas} (2-\gamma)/(2\pi r_c^2) $ is the normalization, 
where $M_{\rm gas}$ is the total gas mass.  The density structure, $\rho(r,z)$, 
is either parametically defined (\S\ref{sec:pedagogic}) or calculated from the 
temperature structure using the equation of hydrostatic equilibrium 
(\S\ref{sec:hydro}).  To calculate the CO number density, we assume that 80\% of
the gas (by number; $\mu = 2.37$) is composed of molecular hydrogen and that the
CO:H$_2$ relative abundance is a constant, $f_{\rm co}$.  Molecular depletion 
due to freezeout is implemented by decreasing $f_{\rm co}$ by a factor of 
10$^{8}$ in the region of the disk where gas temperatures fall below a freezeout
temperature, $T_{\rm gas} < T_{\rm frz}$.  A photodissociation boundary is 
calculated by vertically integrating the H nuclei density 
\citep[$0.706n_{\rm gas}$, the same procedure as in \citet{qi11};][]{aikawa99} and thresholding for 
heights, $z_{\rm phot}$, where the column density is less than a constant value,
\begin{equation}
\sigma_s > 0.706 \int_{z_{\rm phot}}^\infty n_{\rm gas}(r,z') dz'.
\end{equation}
This surface density is equivalent to the unitless $\Sigma_{21}$ defined by
\citet{aikawa06} and used by \citet{qi11} where 
$\Sigma_{21} = \sigma_s / 1.59 \times 10^{21}$\,cm$^{-2}$.

We assume a constant turbulent velocity width, $\xi=0.01$\,km s$^{-1}$, and disk
inclination, $i=224^\circ$.  The bulk gas velocities are described in each 
section and we fix several important model parameters: the stellar mass 
\citep[$M_\ast = 2.3$\,M$_\odot$;][]{natta04}, source distance 
\citep[$d=122$\,pc;][]{vandenancker98}, and major axis position angle 
(measured East of North, PA = 312$^\circ$).  Note that the asymmetries in the 
$^{12}$CO $J$=3$-$2 line emission effectively resolve the absolute disk spin 
orientation \citep{pietu07}, so our PA and inclinations differ by 
$\sim 180^\circ$ from the models presented by \citet{isella07} and 
\citet{qi11}.  See \S\ref{sec:discussion} for a more detailed discussion.

For any given disk structure model, we calculate the CO level populations using 
the non-LTE molecular excitation and radiative transfer package LIME 
\citep{brinch10} and molecular data from the LAMDA database 
\citep{schoier05,yang10}.  LIME is then used to generate channel maps matching 
the native channel spacing of the ALMA data (see Table \ref{tab:lines}).  We
then smooth using the Hanning algorithm in the spectral dimension with a three 
channel kernel to mimic the ALMA pipeline \citep[see 
\S\ref{sec:discussion};][]{lundgren12} and calculate model visibilities to match
the Fourier sampling of the measured visibilities.  We evaluate the model fit 
through a $\chi^2$ statistic of the complex visibilities and also by inspection 
of CLEANed channel maps using the same masks and imaging procedure applied to 
the data (see \S\ref{sec:observations}).

It is important to emphasize that the models developed here involve many 
degenerate parameters: the exact structures we present are neither unique nor 
necessarily optimal.  The analysis presented here is only meant to provide a 
basic physical understanding of some new, key aspects of these data.

\subsubsection{A Pedagogical Structure}\label{sec:pedagogic}

In this section, we will demonstrate how a vertical temperature gradient 
naturally produces the asymmetric emission morphology of the $^{12}$CO $J$=3$-$2
line.  In order to disambiguate the effects of density and temperature, we 
define a single parametric disk density structure and calculate the line 
emission for two simple cases for the temperature structure:
(a) the disk is vertically isothermal and (b) the disk has a 
warm ($T\sim30$\,K) molecular layer above a cold ($T\sim20$\,K) midplane.

\begin{figure}[t!]
\epsscale{0.95}
\plottwo{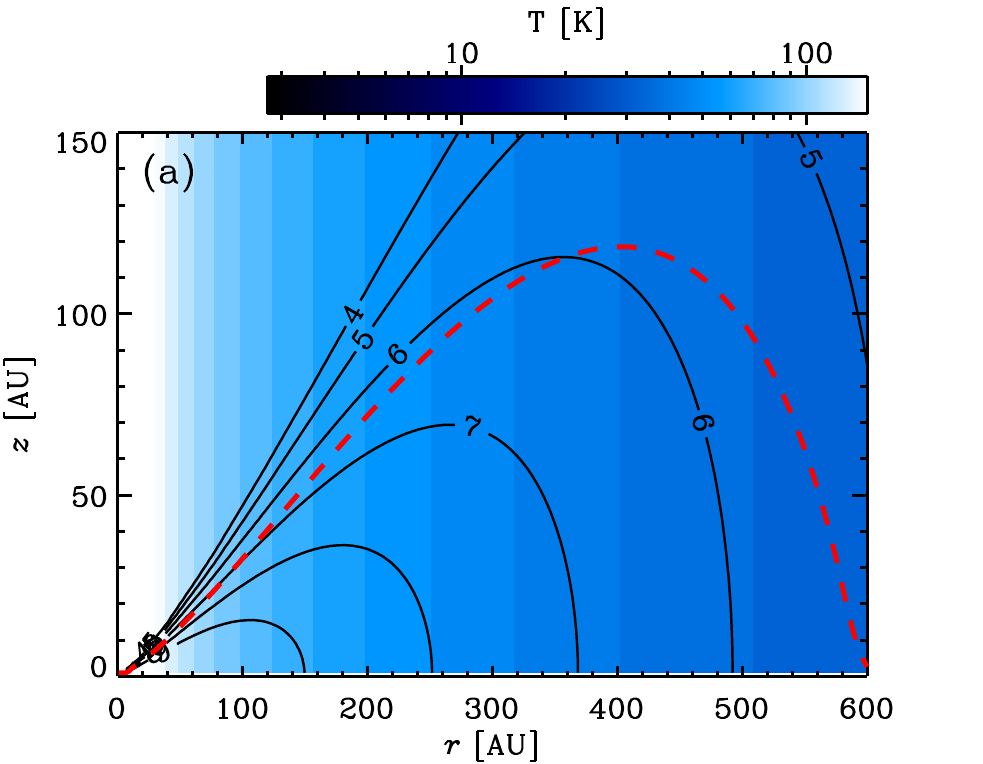}{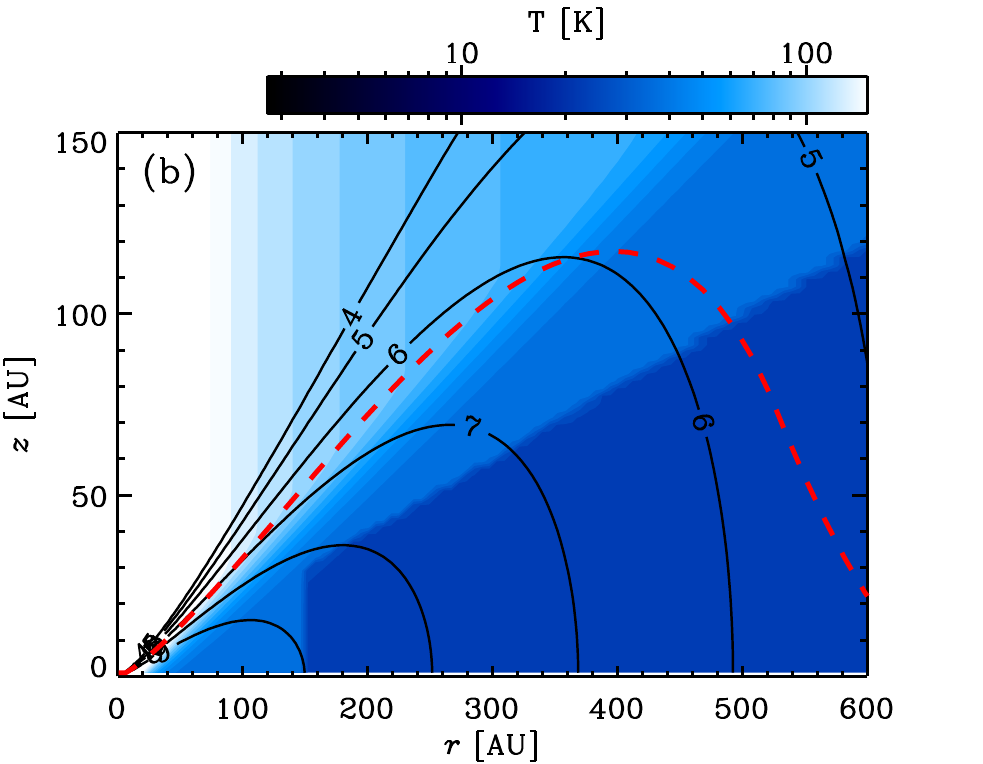}
\figcaption{Temperature ({\it color scale}) and gas density structure 
({\it black contours}; $\log n_{\rm gas} = [4,5,6,7,8,9]$\,cm$^{-3}$) of the 
pedagogical models (\S\ref{sec:pedagogic}).  (a) A vertically isothermal model 
structure. (b) A model structure that features an artificial vertical 
temperature gradient.  The dashed red curves mark the photodissociation 
boundaries, for $\sigma_s = 5\times 10^{-20}$\,cm$^{-2}$.
\label{fig:pedagogical_struct}}
\end{figure}

We assume that the gas is vertically distributed with the Gaussian profile 
appropriate for a vertically isothermal disk in hydrostatic equilibrium 
(neglecting self gravity):
\begin{equation}
\rho_{\rm gas}(r,z) = \frac{\Sigma_{\rm gas}(r)}{\sqrt{2\pi}H} \exp \left[-\frac{z^2}{2H^2}\right].
\end{equation}
For this pedagogical example, the temperature and density structure are 
entirely decoupled. The disk scale height is defined to be 
$H(r) = 16(r/150$\,AU$)^{1.35}$\,AU.  The velocity field is defined such that 
the disk is in circular Keplerian rotation about the central star, with an azimuthal
component
\begin{equation}
{v_K}^2 = \frac{G M_\ast}{r}.
\end{equation}
Depletion due to CO freezeout is not considered in this calculation, but 
photodissociation is included with a threshold density 
$\sigma_s=5\times10^{20}$\,cm$^{-2}$.  The model disk has a gas mass of 
$0.09$\,M$_\odot$, $r_c = 115$\, AU, $\gamma = 0.8$, and 
$f_{\rm co} = 5\times10^{-5}$.

\begin{figure}[t!]
\epsscale{0.95}
\plotone{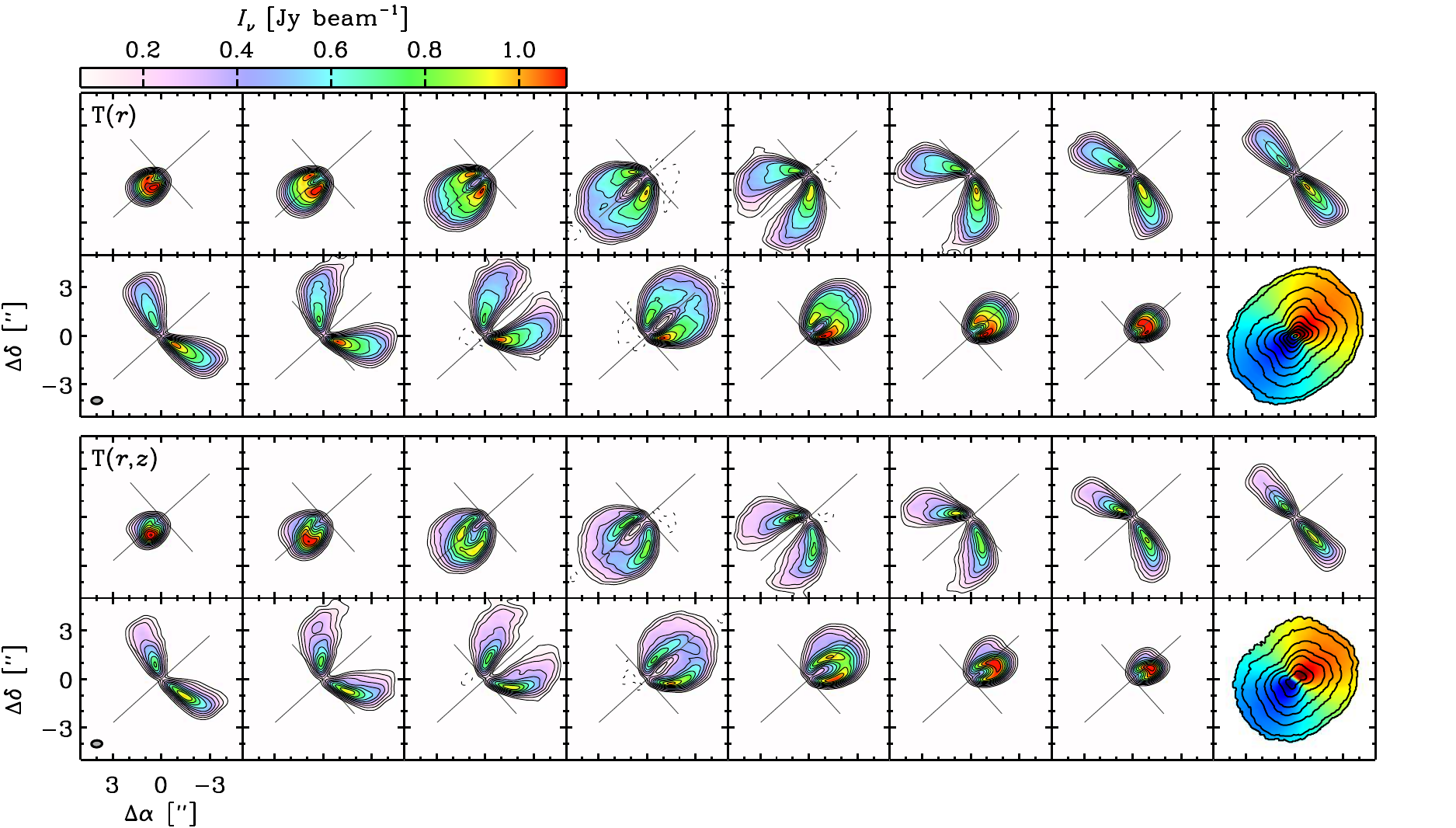}
\figcaption{The $^{12}$CO $J=3-2$ line emission predicted by the models 
described in \S \ref{sec:pedagogic}.  The top set of channel maps shows the 
emission from a vertically isothermal model structure, while the bottom model 
structure has a vertical temperature gradient (see Figure 
\ref{fig:pedagogical_struct}). 
\label{fig:pedagogical}}
\end{figure}

We first calculate the $^{12}$CO $J$=3$-$2 line emission from this disk 
structure assuming that it is vertically isothermal with a radial temperature 
structure set by a power law, $T_{\rm gas} = 65$K\,$(r/100$\,AU$)^{-0.5}$.  The 
disk density and temperature structure is shown in Figure 
\ref{fig:pedagogical_struct}(a), and the corresponding channel maps in Figure 
\ref{fig:pedagogical} (top panels).  The qualitative discussion from 
\S\ref{sec:toy} is confirmed: the morphology of the line is essentially 
symmetric across the major axis with only a slight brightness asymmetry between 
the Northern (near) and Southern (far) sides of the disk (which is reversed if 
the inclination is flipped by $90^\circ$).  This geometric effect, for 
sensitive data, indicates the absolute orientation of the disk and can
qualitatively be understood by considering the emission that originates above 
the disk midplane.  The half of the disk (located in the N) that is tilted 
toward the observer will then have a smaller projection on the observer sky 
plane (and appear dimmer) than the disk half (located in the S) that is tilted 
away.  The double cone structure described in \S\ref{sec:toy} similarly
encapsulated this effect: in Figure \ref{fig:cone_chmaps} the isovelocity 
contours of the near cone (black curve) appear noticeably shorter in the N part
of the disk than in the S for channels near the line center, $|v_{\rm los}| 
\lesssim 1.02$\, km s$^{-1}$.  Optical depth effects and the disk
structure also contribute to the details of this brightness asymmetry.

Next, we build on that calculation by including a vertical gradient to the
temperature structure.  We start by defining a ``warm'' region where the gas is 
a constant $T_ a$ = 30\,K for $z > 20$\,AU\,$(r/100$\,AU) or $r < 150$\,AU 
\citep[the freezeout radius for this disk][]{qi11}.  Outside this 
region (i.e., the disk midplane for $r > 150$\,AU), the gas is a cooler 
$T_ m$ = 20\,K.  Since the gas high above the midplane and close to the star
should be hotter than the 30\,K typical for the outer region of this disk,
we add a hot atmosphere above $z_a = 73$AU\,$(r/200$\,AU$)$ with a power-law 
component $T_a(r,z>z_a) = T_{\rm gas} + $40K\,$(r/200$\,AU$)^{-0.8}$.  
This disk structure is shown in Figure \ref{fig:pedagogical_struct}(b), and the 
corresponding channel maps in Figure \ref{fig:pedagogical} (bottom panels).  
Qualitatively, this model does an excellent job of producing both the 
morphological and brightness asymmetries noted in \S\ref{sec:results}.  

The disk structures described above are both massive and geometrically thick: 
the model scale height $H$ is much larger than what the actual midplane 
temperatures would produce.  Specifically, the defined $H(r)$ would correspond 
to a midplane temperature of $T_m = 50$K\,$(r/100$\,AU$)^{-0.3}$, which is too 
warm to reproduce the line asymmetries (too much emission would be generated, 
essentially filling in the hollow cone emission morphology).  However, the 
artificially large scale height is necessary to reproduce the morphology, since 
increasing the height of the emitting layer increases the projected separation 
of the front and back half of the disk (see \S\ref{sec:toy}).  This suggests 
that a warm layer suspended above a cold midplane may help to raise the height 
of the emitting CO layer.  It should be emphasized that a vertical temperature 
gradient is necessary to reproduce the data although it may not be the only 
contributing effect.  Depletion of the gas phase CO due to freezeout onto grains
should also reduce the contribution of the midplane emission, but this effect 
requires the cold temperatures that, by themselves, can produce this 
distinctive morphology.

\subsubsection{A Hydrostatic Model Structure}\label{sec:hydro}
While the previous model serves to demonstrate how a vertical temperature 
profile can explain the observed asymmetries in the $^{12}$CO $J$=3$-$2 and 
$^{12}$CO $J$=2$-$1 lines, the model structure is purely pedagogical and not 
physically self-consistent.  We now present a hydrostatic disk model where the 
densities are internally consistent with the parametrically defined 
temperatures.  This model roughly reproduces both the morphology and intensity 
of the observed emission simultaneously for the lines of interest, 
$^{12}$CO $J$=3$-$2, $^{12}$CO $J$=2$-$1, $^{13}$CO $J$=2$-$1, and 
C$^{18}$O $J$=2$-$1.

\begin{figure}[t!]
\epsscale{0.45}
\plotone{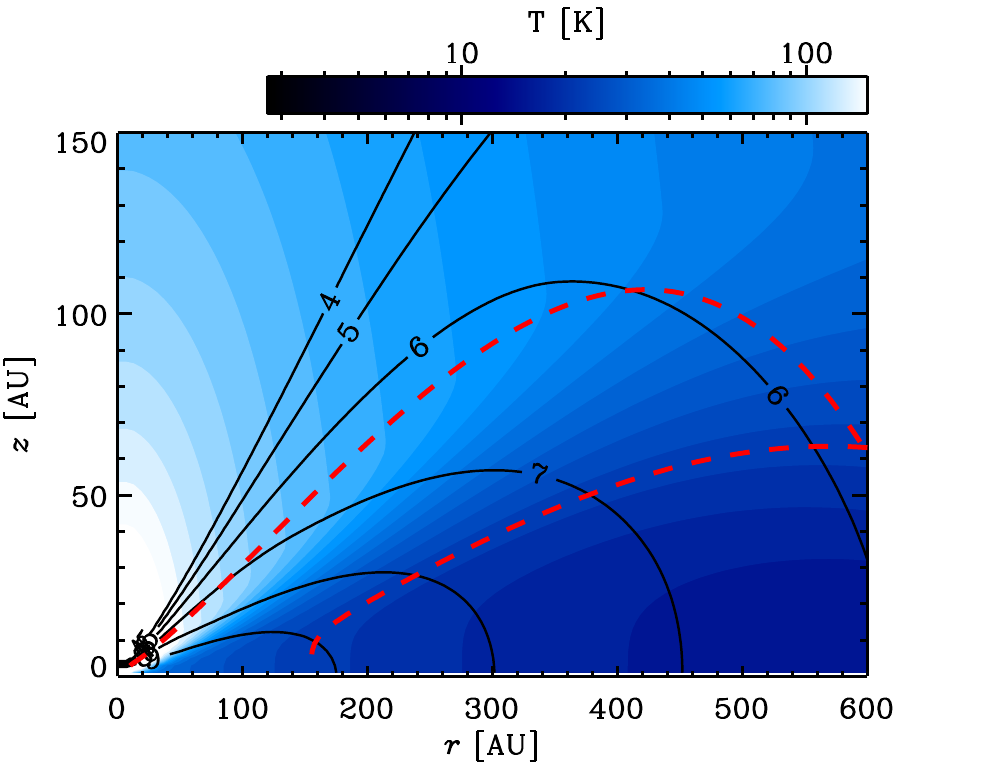}
\figcaption{The temperature and density structure of the hydrostatic model 
described in \S\ref{sec:hydro}, using the same legend as Figure 
\ref{fig:pedagogical_struct}.  The lower red curve indicates the upper boundary 
of the cold midplane where the gas phase CO densities are reduced due to 
freezeout onto grains.
\label{fig:hydro_struct}}
\end{figure}

\begin{figure}[t!]
\epsscale{0.95}
\plotone{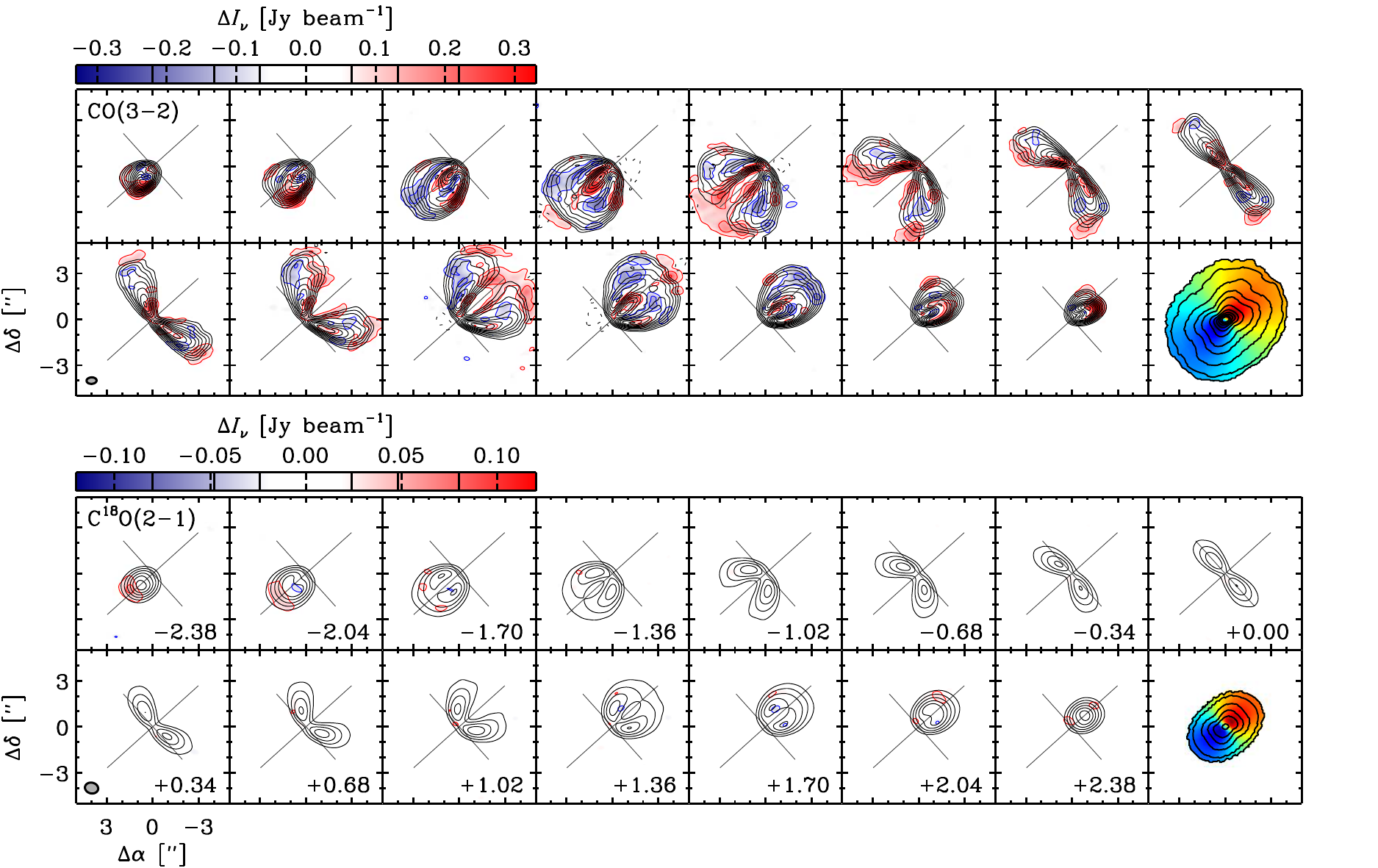}
\figcaption{The $^{12}$CO $J$=3$-$2 ({\it top}) and C$^{18}$O $J$=2$-$1 line 
emission ({\it bottom}) of the hydrostatic model described in \S\ref{sec:hydro}.
For both lines the imaged residual visibilities ($\Delta I_\nu=$ data -- model) 
are shown in color ({\it blue} is negative and {\it red} is positive). The 3, 6,
and 10\,$\sigma$ levels are indicated by vertical black lines in the colorbar.
\label{fig:hydrochmaps}}
\end{figure}

\begin{figure}[t!]
\epsscale{0.95}
\plotone{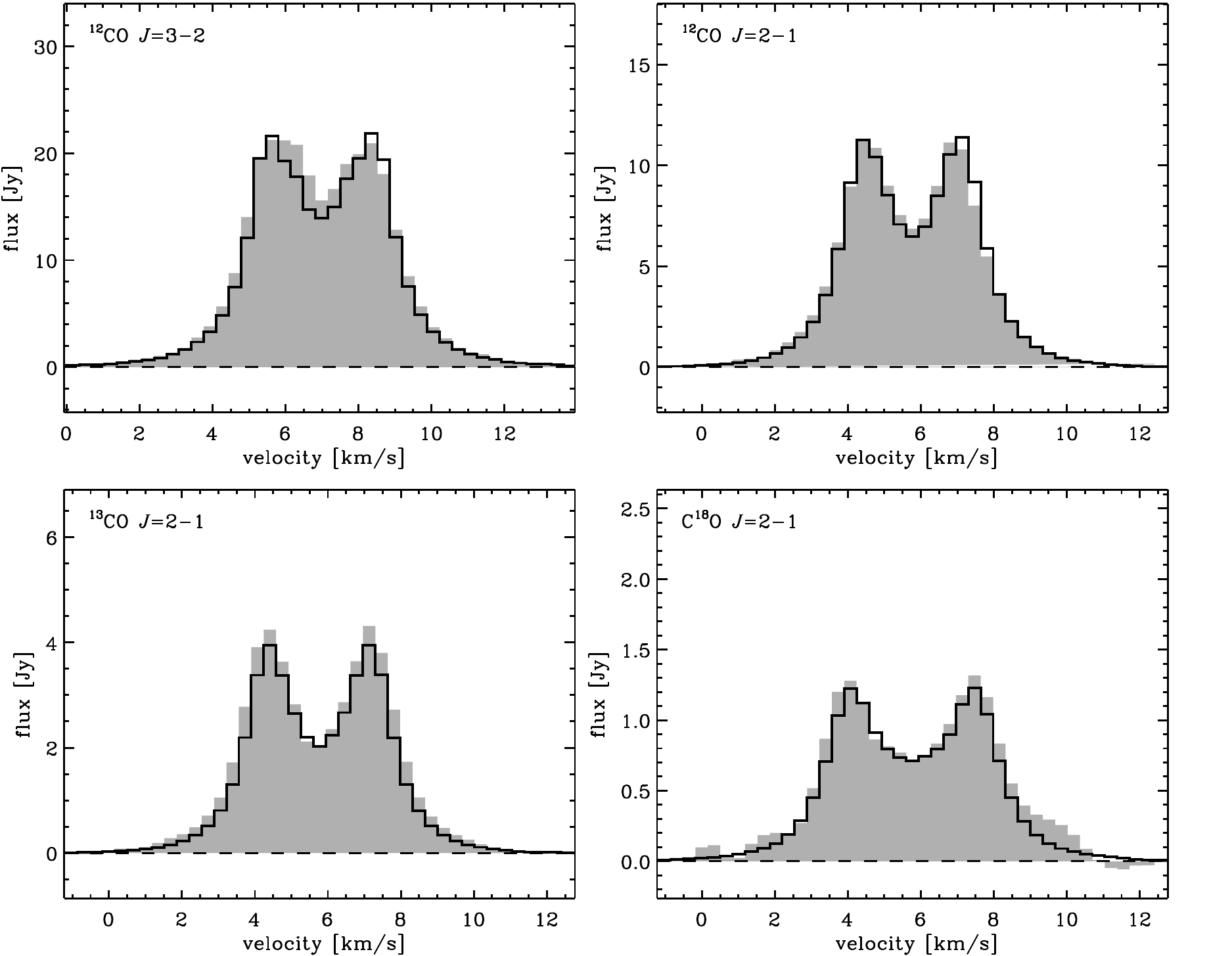}
\figcaption{The integrated line profiles of the four observed CO lines 
({\it gray}) compared to the model spectrum ({\it black lines}; 
\S\ref{sec:hydro}).  The spectra were calculated in 10\arcsec\,square boxes 
except for the C$^{18}$O line, which used a 6\arcsec\,square box.
\label{fig:hydrolps}}
\end{figure}

We define the gas temperatures using a slightly modified version of the 
parameterization introduced by \citet{dartois03}, where
\begin{equation}
 T_{\rm gas}(r, z) = \left\{
\begin{array}{ll}
T_a + \left(T_m - T_a\right) \left[\sin{\frac{\pi z}{2 z_q}}\right]^{2\delta(r)} & \mbox{if $z < z_q$} \\
T_a & \mbox{if $z \ge z_q$} 
\end{array}.
\right. 
\label{eq:dartoisII}
\end{equation}
Here, $T_m(r,z=0) = 19$K\,$(r/155$\,AU$)^{-0.3}$ is the temperature at the disk 
midplane and the atmosphere temperature above the height $z_q$ is 
\begin{equation}
T_a(r, z) = 55{\rm K\,}(\sqrt{r^2 + z^2}/200{\rm \,AU})^{-0.5}.
\end{equation}
This prescription produces a smooth, monotonically increasing (with $z$) 
transition between $T_m(r,z=0)$ and $T_a(r,z=z_q)$, with the vertical shape 
controlled by the parameter $\delta = 0.0034*(r-200$\,AU$) + 2.5$ (with the 
requirement that $\delta \geq 0.3$).  The height of the disk atmosphere is 
assumed to have a radial distribution described by a truncated power law, 
$z_q = 63$AU\,$(r/200$\,AU$)^{1.3} \exp[-(r/800$\,AU$)^2]$.  The midplane 
temperature profile we assume has the same CO freezout radius calculated by 
\citet{qi11}, but with a shallower power-law index.  The form of the vertical 
temperature profile (Equation \ref{eq:dartoisII}) was designed to mimic the 
dust temperatures in the disk models developed by 
\citet{dalessio98,dalessio99,dalessio01,dalessio06}.  Since we do not do the 
radiative transfer calculation for a dust disk 
\citep[e.g.,][]{qi11,andrews12}, we adopt this parameterization as a natural 
choice for the outer regions of a disk.

The vertical distribution of the gas is determined by solving the differential 
equation for hydrostatic equilibrium with our parametrically defined temperature
structure:
\begin{equation}
-\frac{\partial \ln \rho_{\rm gas}}{\partial z} =\frac{\partial\ln T_{\rm gas}}{\partial z} + \frac{1}{{c_s}^2}\left[\frac{G M_\ast z}{(r^2 + z^2)^{3/2}}\right],
\end{equation}
where ${c_s}^2 = k_B T_{\rm gas}/\mu m_h$ is the sound speed.  We normalize the 
density so that the total mass of the gas disk is 
$M_{\rm gas} = 0.09$\,M$_\odot$, and assume $\gamma = 0.8$ and $r_c=150$\,AU.  
As before,  we assume that the gas follows circular Keplerian orbits and that the 
self-gravity of the disk is negligible.   We now include freezeout for 
$T_{\rm frz} = 19$\,K and photodissociation with 
$\sigma_s = 9\times10^{20}$\,cm$^{2}$.  For the CO isotopes, we assume the 
relative abundances measured for the ISM by \citet{wilson99}: 
$^{12}$CO/$^{13}$CO$=69\pm6$ and $^{12}$CO/C$^{18}$O$=557\pm30$.  The disk 
density and temperature structure are shown in Figure \ref{fig:hydro_struct}.

In Figure \ref{fig:hydrochmaps} we present channel maps of predicted model 
$^{12}$CO $J$=3$-$2 and C$^{18}$O $J$=2$-$1 emission, with the 
imaged residuals overlaid in color.  This model successfully reproduces the 
distinctive spatial morphology of the $^{12}$CO $J$=3$-$2 line (top panels) 
while also appearing symmetric in the C$^{18}$O $J$=2$-$1 line (bottom panels). 
The majority of the C$^{18}$O emission is contributed from material
much closer to the midplane than in the case of $^{12}$CO.  The resulting 
signature of the temperature asymmetry is much less pronounced in the 
isotopologue lines, and is unresolved by these data.  We show a comparison of 
the integrated line profiles for the data and model predictions in Figure 
\ref{fig:hydrolps} for the four CO lines of interest. 

While this model matches the data fairly well, there are systematic residuals at
large radii near the major axis at projected velocities of $\approx 1$\,km
s$^{-1}$ where the model does not produce any significant emission 
$(I_\nu < 3\sigma)$.  We cannot account for this emission by increasing the 
disk size ($r_c$), since emission at larger radii for channels at 
higher velocities as well as at the systemic velocity is overproduced.  One 
interpretation of these significant residuals is that the gas at large radii is 
predicted to be moving {\it too fast} and so the emission morphology of the 
model appears to ``fold'' toward the major axis too slowly (when starting at 
the line center and looking channel by channel toward the blue or red shifted 
line wings).  We address how the model may not accurately describe the true disk
velocity field in the next section.

\subsection{A Closer Look at the Gas Velocities}
\label{sec:velocity}
Although the models we present in \S\ref{sec:pedagogic} and \S\ref{sec:hydro} 
leave non-negligible residuals compared to the data, they nevertheless serve to 
illustrate how a disk with a realistic vertical temperature structure naturally 
explains the asymmetric morphology of the $^{12}$CO $J$=3$-$2 line.  These data 
are an example of the exquisite sensitivity that ALMA will routinely achieve for
observations of protoplanetary disks.  With this in mind, we now explore how 
these data are sensitive to the bulk motions of the gas in the HD 163296 disk, 
necessitating the use of self-consistent, physically motivated disk structures 
in any detailed analysis of its structure or kinematics.

We begin with a broad discussion of the basic expectations for gas velocities in
protoplanetary disks.  These disks are usually assumed to be rotating 
differentially and in vertical hydrostatic equilibrium, so that there is no bulk
vertical motion of the gas.  At any location in the disk, there is an orbital 
velocity,
\begin{equation}
\vec{v} = v \,\hat{\theta}
\end{equation}
where $\hat{\theta}=(\sin\theta, -\cos\theta,0)$, and the gas is assumed to be 
on circular orbits.  The standard \citep[or tested, e.g.][]{dutrey94} assumption
is that $v$ is equivalent to the Keplerian orbital velocity set by the 
gravitational potential of the central star,
\begin{equation}
\frac{{v_K}^2}{r} = \frac{G M_\ast}{r^2}.
\label{eq:fkep}
\end{equation}
Equation \ref{eq:fkep} implicitly assumes that the disk is geometrically thin, 
so that for a given $r$ the gas at the midplane ($z=0$) is moving at the same 
velocity as the gas suspended above it.  But including the vertical geometry 
actually decreases the force from the star as well as the radial projection of 
that force vector (which sets the orbital velocity).  This effectively reduces 
$v$ as $z$ increases, and implies that the disk has differential rotation in 
both the radial and vertical dimensions,
\begin{equation}
\frac{v^2}{r} = \frac{r}{(r^2 + z^2)^{1/2}} \left(\frac{G M_\ast}{r^2 + z^2}\right).
\label{eq:fgeom}
\end{equation}
Recasting in terms of the toy model described in \S\ref{sec:toy}, Equation
\ref{eq:fgeom} reduces the observed velocity of the CO emitting layer by a 
factor of $(\cos\psi)^{3/2} \sim 0.95$--$0.91$ for a cone angle, $\psi$, of 
15--20$^{\circ}$.  This rough estimate alone suggests caution when adopting a 
Keplerian, thin disk velocity field (Equation \ref{eq:fkep}) for modeling 
emission from molecules located high above the midplane.  In principle, this
assumption can be checked {\it a posteriori} for any disk model by calculating 
$\psi$ (or equivalently $z$) of the $\tau=1$ emitting surface 
(see \S\ref{sec:discussion}).

In addition to the gravitational potential of the central star, any radial 
change in the gas pressure will provide a force term and alter the gas 
velocities from the fiducial Keplerian field, $v_K$.  Including the pressure 
gradient, the force equation becomes
\begin{equation}
\frac{v^2}{r} = \frac{G M_\ast r}{(r^2 + z^2)^{3/2}} + \frac{1}{\rho_{\rm gas}}\frac{\partial P_{\rm gas}}{\partial r}.
\end{equation} 
As both the gas temperature and density tend to decrease with $r$, a 
corresponding negative pressure gradient causes the gas to slow down and orbit 
at sub-Keplerian velocities.  This phenomenon is thought to play an important 
role in the radial migration and growth of solids in these disks 
\citep{weidenschilling77b,takeuchi02,birnstiel10}, but is a subtle effect that 
slows down the gas from the fiducial Keplerian velocities by a rough factor of 
$(1-c_s^2/v_K^2)^{1/2} \sim 0.99$ for a temperature of 30\,K at $r\sim 500$\,AU 
in the disk \citep{armitage09}.  However, if the density is falling faster with 
radius than a power-law, the pressure gradient will be larger and the gas 
velocities markedly slower.  Therefore, the significance of this force term 
depends intimately upon the density and temperature structure.  The popular 
self-similar solution for $\Sigma_{\rm gas}(r)$ that we have utilized features 
an exponential tail that will enhance this effect.  

The last force term we consider is the self-gravity of the disk.  Unlike the two
previous terms which tended to slow down the gas and produce sub-Keplerian 
velocities, the additional mass contributed by the disk should increase $v$.  
The composite force equation is then
\begin{equation}
\frac{v^2}{r} = \frac{G M_\ast r}{(r^2 + z^2)^{3/2}} + \frac{1}{\rho_{\rm gas}}\frac{\partial P_{\rm gas}}{\partial r} + \frac{\partial \phi_{\rm gas}}{\partial r},
\label{eq:fullF}
\end{equation} 
where $\phi_{\rm gas}$ is the potential due to self-gravity of the disk.  The 
right hand side of this equation includes three force terms: 
$F_{\rm stellar\,gravity}$, $F_{\rm pressure\,gradient}$, and 
$F_{\rm disk\,gravity}$ (appearing from left to right in Equation 
\ref{eq:fullF}).  To self-consistently construct a disk structure now requires 
iteratively solving the equation of hydrostatic equilibrium,
\begin{equation}
-\frac{\partial \ln \rho_{\rm gas}}{\partial z} = \frac{\partial\ln T_{\rm gas}}{\partial z} +  \frac{1}{{c_s}^2}\left[\frac{G M_\ast z}{(r^2 + z^2)^{3/2}} + 
\frac{\partial \phi_{\rm gas}}{\partial z}\right],
\end{equation}
and Poisson equation,
\begin{equation}
\nabla^2 \phi_{\rm gas} = 4 \pi G \rho_{\rm gas}.
\end{equation}
For most disk configurations, the Poisson equation must be calculated 
numerically which we do using a fixed grid \citep{swarztrauber75} 
{\it without} any iterations with the hydrostatic equilibrium equation (i.e., we
do not include self-gravity when calculating the disk density structure for 
reasons explained below).  As an illustration, we can roughly estimate the
contribution of this correction by considering a special case with an analytic 
solution to the Poisson equation: a thin disk with a surface density profile 
$\Sigma_{\rm gas} = \Sigma_0 r_0 /r$ and no outer edge has 
\begin{equation}
\left.\frac{\partial \phi_{\rm gas}}{\partial r}\right|_{r=R} = 2\pi G\Sigma(R)
\end{equation}
at the midplane \citep{mestel63,lodato07}.  In that scenario, the gas velocities
are increased by a factor of 
$(1+ 2\pi Gr\Sigma(r)/ v_{\rm Kep}^2)^{1/2} \sim 1.005$--$1.001$ for radii 
$\sim 100$--$500$\,AU in the disk.  We note that throughout our analysis we have
only considered circular orbits.  Including eccentricity in the disk 
\citep[e.g.,][]{regaly11} introduces a host of new parameters, asymmetries in the
integrated line profile (which we have not investigated), and is beyond the 
scope of our analysis.

\begin{figure}[t!]
\epsscale{0.95}
\plotone{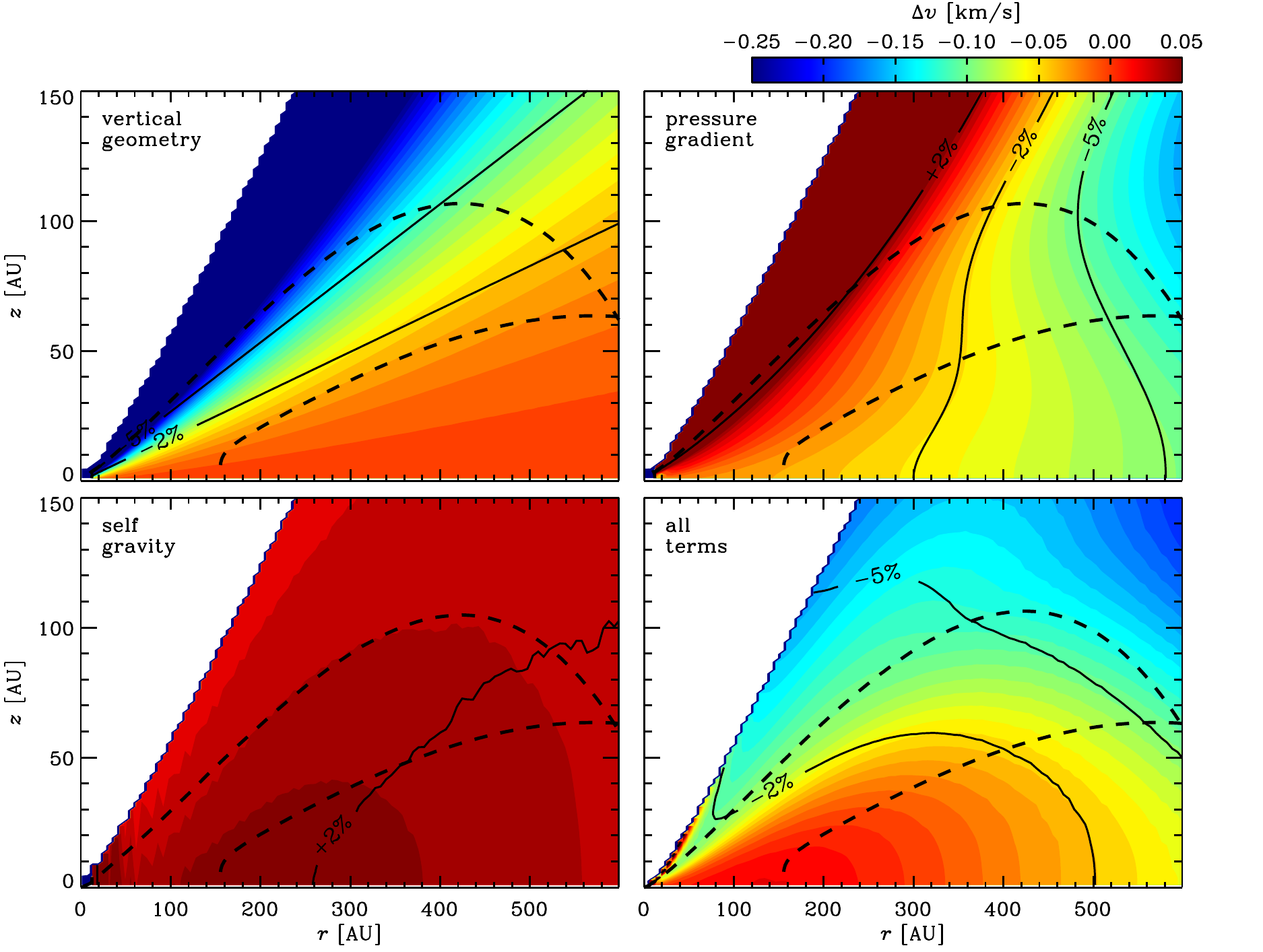}
\figcaption{The velocity difference ({\it in color}) between the thin Keplerian 
disk (${v_K}^2 = G M_\ast /r$) and the four models
considered here.  The black contours mark the 
fractional difference (in \%).  For reference, the native channel width of
the $^{12}$CO $J$=3$-$2 data is $0.11$\,km s$^{-1}$. The molecular layer of the 
model (\S\ref{sec:models}) is indicated by the dashed black border.
\label{fig:veloc}}
\end{figure}

We now explore whether these ALMA data can distinguish between models that 
include these subtle effects on the gas velocities: (1) the differential 
rotation due to the vertical geometry of the disk, (2) the sub-Keplerian 
velocities of the gas due to the radial pressure gradient, and (3) the 
super-Keplerian velocities introduced by self-gravity.  In order to disambiguate
these three effects, we will consider separately the three force terms in 
Equation \ref{eq:fullF} that determine $v$: $F_{\rm stellar\,gravity}$, 
$F_{\rm pressure\,gradient}$, and $F_{\rm disk\,gravity}$.
When modeling the last two disk-specific terms, we revert to the thin-disk 
approximation where $F_{\rm stellar\,gravity}=GM_\ast/r^2$.  We use the same 
observing geometry, disk density and temperature structure introduced in 
\S\ref{sec:hydro}, changing only the gas velocities.  As mentioned
previously, the self gravity of the disk affects both the velocity and 
density structures of the disk.  Since both of these structures will impact the 
observed line emission and we are interested here in the effect of the former 
only, we choose to omit the disk potential in the equation of hydrostatic 
equilibrium and leave the disk density structure unaltered.  For completeness, 
we also calculate the emission from a model whose velocities are determined by 
all three terms (Equation \ref{eq:fullF}).  Due to the more complicated velocity
field and the fine spatial sampling of our disk models, we found that the 
packaged raytracer in LIME was prohibitively slow.  Therefore, for these models 
we only used LIME to calculate the non-LTE level populations.  To generate the 
model images, we utilized the axisymmetry of our disk models to interpolate 
onto a fine 2D grid \citep{fan05} and integrated the radiative transfer 
equation.  None of these four models require any additional parameters.

\begin{figure}[t!]
\epsscale{0.86}
\plotone{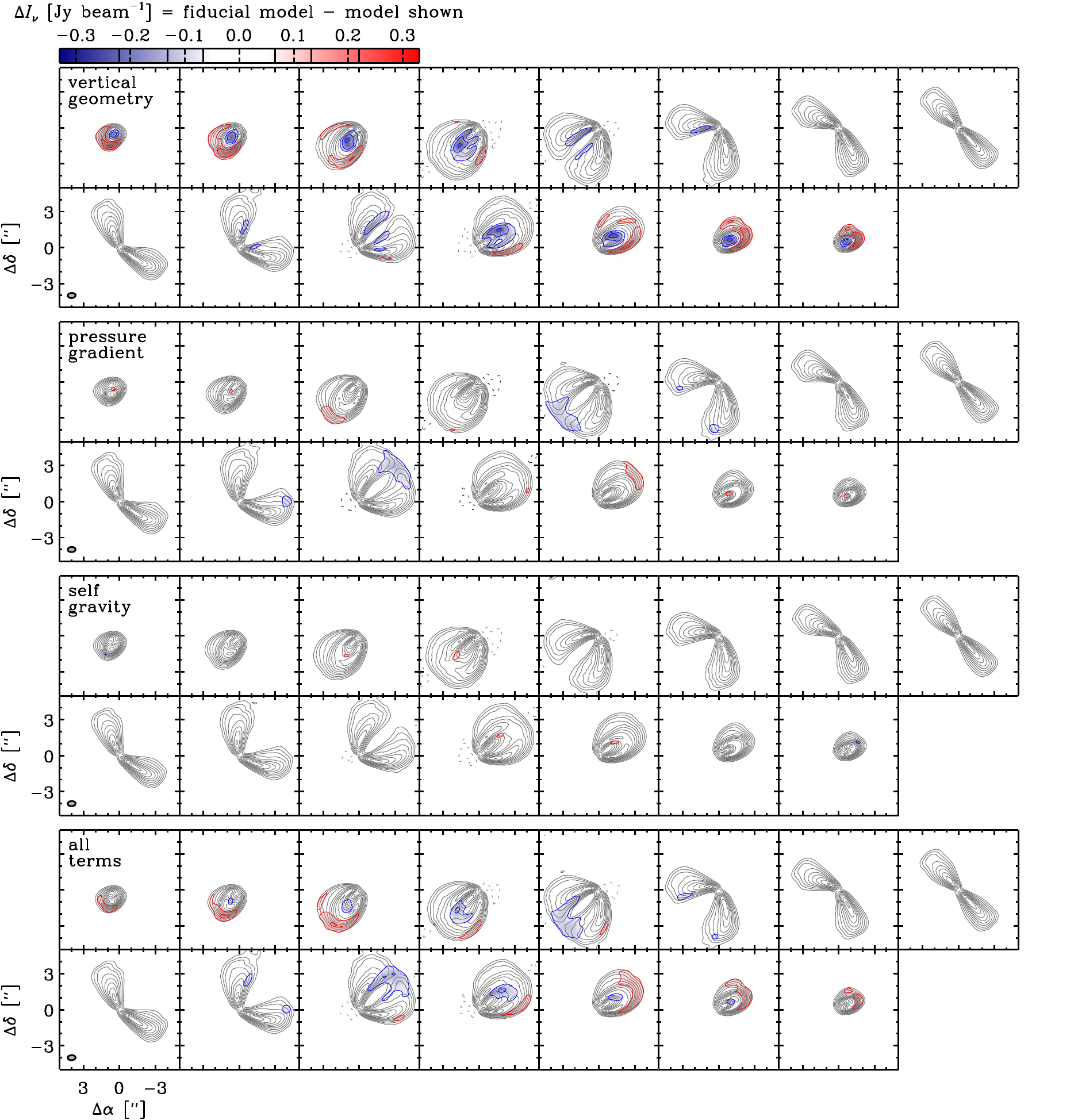}
\figcaption{The synthesized channel maps of the models (shown in {\it gray}) 
with the various velocity structures described in \S\ref{sec:velocity}.  
Overplotted in color are the fiducial model $-$ model residuals.  The fiducial 
model (\S\ref{sec:hydro}) assumes that the disk is thin ($z \ll H$) with 
Keplerian orbits ($v^2 = G M_\ast/r$; $r=\sqrt{x^2 + y^2}$).  The residuals do 
{\it not} indicate the quality of the model as a fit to the data: they show how 
the model emission changes when the velocity field is changed.
\label{fig:effects}}
\end{figure}

Figure \ref{fig:veloc} shows the fractional and absolute velocity difference 
compared to the fiducial Keplerian velocity field (${v_K}^2 =G M_\ast/r$) for 
each model structure. As expected, the geometry of the disk height and the disk 
pressure gradient slow down the gas high above the midplane and at large disk 
radii ($r \gtrsim 300$\,AU).  The disk self-gravity speeds up the gas, but has 
a smaller effect than the first two terms.  Combining all three effects 
for our fiducial model, the fractional difference in the gas velocities are on 
the order of a few percent in the midplane.  However, the absolute velocity 
difference in the CO emitting region in the outer disk can be as great as 
$\sim0.1\,$km s$^{-1}$, the same as the native channel width of the Band 7 
observations. 

The impact of these changes to the velocity field are significant at the 
$\gtrsim 3 \sigma$ level across many channels.  Figure \ref{fig:effects} shows 
the difference between the model with the standard Keplerian rotation 
($v^2 =G M_\ast/r$; \S\ref{sec:hydro}) and the individual models whose velocity 
fields are summarized in Figure \ref{fig:veloc}.  For all of these models, the 
central channels were the least affected since the projected velocity (for this 
inclination) is small ($v_{\rm los} \propto \cos \theta$).  The model that 
accounts for the vertical geometry of the disk differs most significantly, 
particularly for the higher velocity channels ($v \gtrsim 1.4$\,km s$^{-1}$).  
The radial pressure gradient impacts the emission at large radii, where the 
surface density profile falls off sharply.  By itself, the disk self gravity has
the least significant effect.  However, the combined model is clearly not 
dominated by any one term, indicating that models for data of this high quality 
need to incorporate all of these more physical and self-consistent calculations 
for the velocity field.  The usual assumptions can, of course, be instead tested
{\it a posteriori} as we have done here.

\section{Discussion}\label{sec:discussion}

We have conducted an analysis of multiple CO emission lines observed toward the
protoplanetary disk hosted by HD 163296.  The exquisite spatial resolution and 
sensitivity of these ALMA SV observations, combined with good spectral 
resolution of the $^{12}$CO $J=$3$-$2 line, clearly reveals that the spatial 
morphology of the emission is asymmetric across the major axis of the disk.  We 
interpret this 
asymmetry as a resolved signature of the vertical temperature gradient in the 
disk generically predicted by physical/chemical models of disk structures 
\citep[e.g.,][]{pavlyuchenkov07,semenov08}.  We developed a series of 
physically self-consistent disk structures, showing that these ALMA data can 
distinguish between models that correct the gas velocities for the vertical 
thickness of the disk and the radial pressure gradient of the gas.

We started with a toy model that can conceptually explain the asymmetric 
morphology of the $^{12}$CO $J$=3$-$2 line.  An emitting surface described by a 
rotating double cone that makes an angle $\psi \sim 15^\circ$ with the disk 
midplane naturally mimicked the observed morphology of this disk.  Such a toy 
model is analogous to a layered disk structure, where the contribution of the 
midplane emission is small due to the cold gas temperatures and depleted 
abundances produced by the freezeout of gas phase CO onto grain surfaces.  We 
confirmed the qualitative behavior of this toy model by calculating the non-LTE 
emission from both pedagogical and hydrostatic axisymmetric disk models, 
concluding with a disk structure that roughly reproduced the observed morphology
of the $^{12}$CO $J$=3$-$2 emission.  Our analysis also showed that a vertical 
temperature gradient, which has been reported in this disk by both 
\citet{qi11} and \citet{akiyama11}, was consistent with the observed 
intensities of the $J$=2$-$1 emission from $^{12}$CO, $^{13}$CO, and C$^{18}$O, 
as well as the symmetric morphology of the rarer isotopologue lines.

Spectral line emission, even when it is spatially resolved, cannot easily define
the absolute orientation of the disk: there is a 180$^\circ$ ambiguity in the 
inclination and position angle.  However, the morphological and brightness 
asymmetry highlighted here resolves the absolute orientation of the disk.  We 
report values of the disk inclination and position angle that differ by 
180$^\circ$ from previous analyses based on less sensitive data 
\citep{isella07,hughes08,qi11}.  For HD 163296, an independent check on the 
viewing geometry is provided by its outflow, where Doppler shift measurements 
of H$\alpha$, [S II], O III], [N II] emission lines and imaging can 
differentiate between the jet (blue shifted, seen on the near side of the disk) 
and counter jet (red shifted, partially hidden behind the disk).  
For HD 163296, the blueshifted jet is oriented in the SW direction while the 
redshifted counter jet is moving NE \citep[in agreement with the disk wind 
observed via $^{12}$CO by][]{klaassen13}.  This is consistent with our reported 
inclination axis, which tilts the NE part of the disk toward the observer.  
Furthermore, the projecion of the disk rotation axis on the observer sky plane 
(42$^\circ$) is aligned with the PA of the jet \citep[measured for the 
counter jet as $42.0 \pm 3.5^\circ$;][]{grady00}.  The 
jet inclination \citep[$51_{-9}^{+11}$$^\circ$;][]{wassell06,gunther13} also 
agrees with the disk inclination within the $1\sigma$ error bars.  We confirm 
that the jet appears to be orthogonal to the disk plane, although the 
large error bars in the jet inclination leaves open the possibility that the
jet is slightly skewed ($\Delta i \lesssim 10^\circ$).

\begin{figure}[t!]
\epsscale{0.95}
\plotone{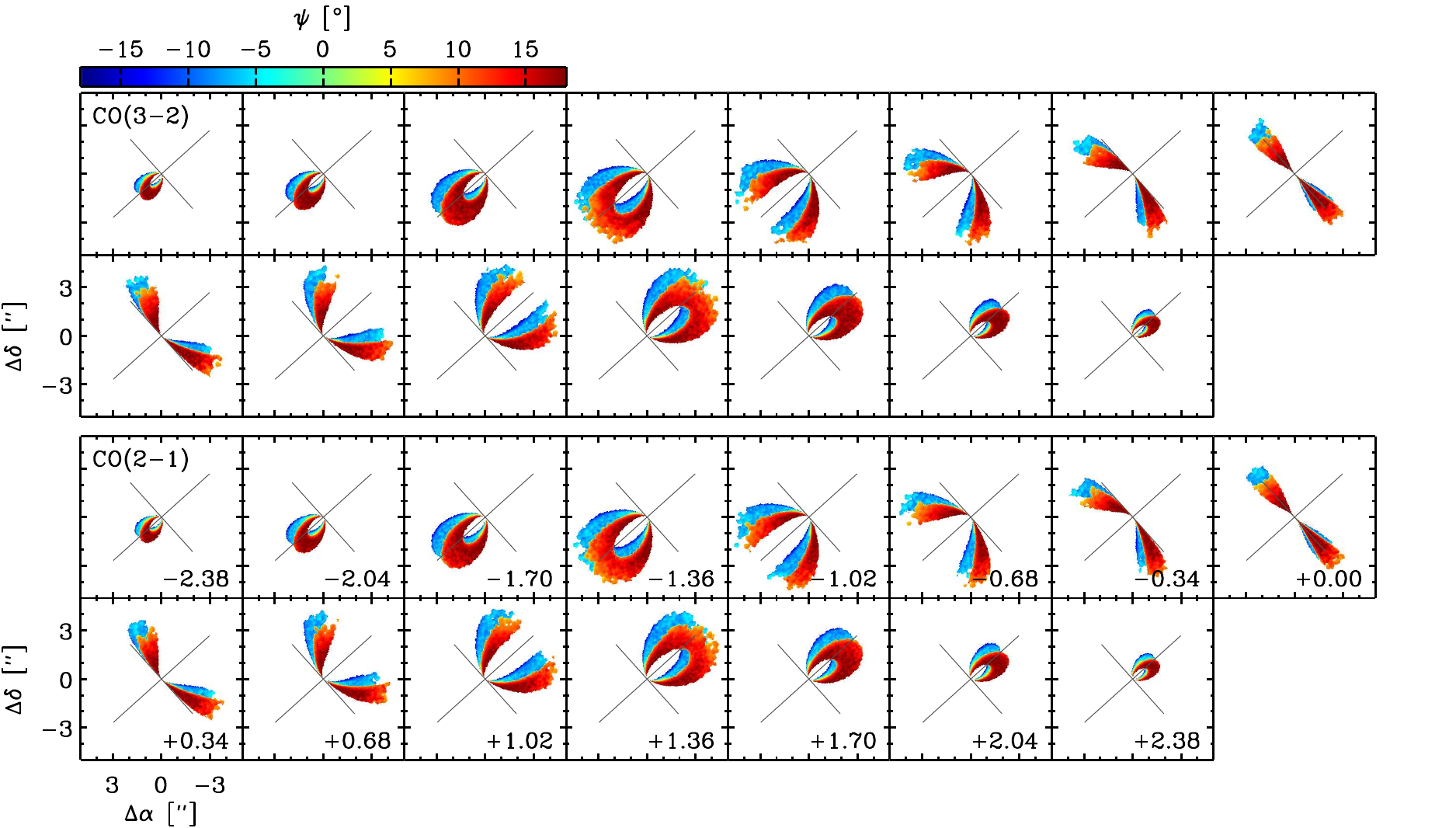}
\figcaption{Map renderings of the angle, $\psi = \tan^{-1}(z/r)$, that the 
$\tau = 1$ surface of the $^{12}$CO $J$=3$-$2 ({\it top}) and $^{12}$CO 
$J$=2$-$1 ({\it bottom}) model emission makes with the disk midplane (for the 
fiducial model described in \S\ref{sec:hydro}). 
\label{fig:psi}}
\end{figure}

\begin{figure}[t!]
\epsscale{0.95}
\plotone{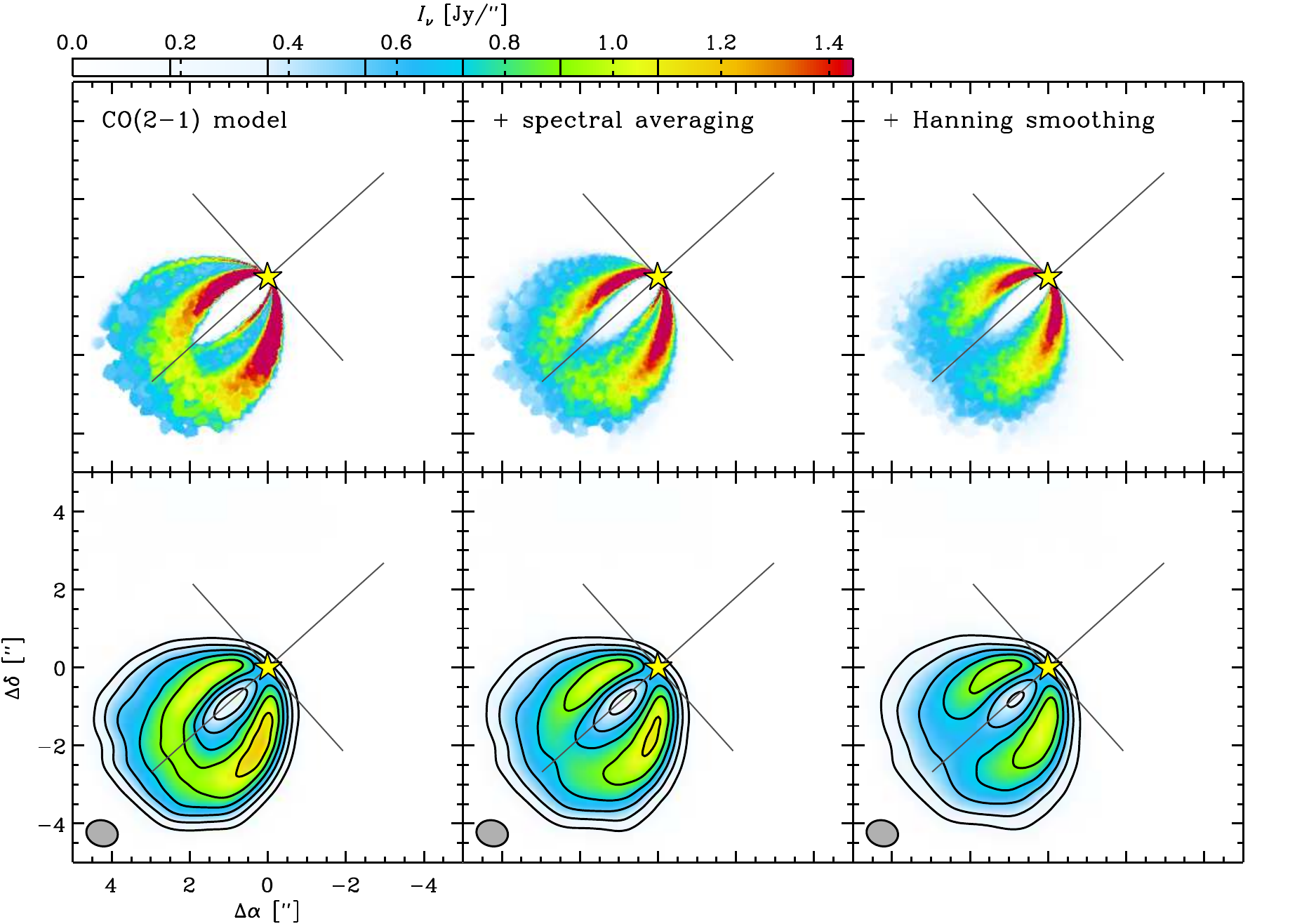}
\figcaption{This set of panels show the effect that the smoothing and averaging 
in the spectral dimension has on the apparent asymmetry of a model.  The top 
row features the high fidelity $^{12}$CO $J$=2$-$1 model observed at 
$v=-1.36$\,km s$^{-1}$ ({\it left}) which is then spectrally averaged over 
a spectral bin width of $\Delta v = 0.32$\,km s$^{-1}$ ({\it center}), and then 
Hanning smoothed ({\it right}). The bottom row shows the synthesized image for 
each model and all panels are shown on the same color scale with
$5\sigma$ interval contours marked on the color bar.
\label{fig:spec_effects}}
\end{figure}

A subtle, but important, point is that spatially resolving the asymmetry caused
by the vertical temperature gradient requires that the data has both excellent 
spatial {\it and spectral} resolution.  In Figure \ref{fig:psi} we show a 
pixel map of the angle, $\psi = \tan^{-1}(z/r)$, that the $\tau=1$ surface of 
our fiducial model (\S\ref{sec:hydro}) makes with the disk midplane for the 
$^{12}$CO $J$=3$-$2 and $^{12}$CO $J$=2$-$1 emission.  Both of these lines 
originate at nearly the same height, and both models exhibit an asymmetric 
emission morphology at a $\Delta v = 0.1$km s$^{-1}$ resolution.  However,
when imaged at the ALMA SV resolutions, the $J$=2$-$1 line appears mostly 
symmetric -- just like the data (exceptfor the slight rotation at 
$|v| \approx 1.7$\,km s$^{-1}$).  Figure \ref{fig:spec_effects} shows the 
accumulated degradation of the spectral 
resolution for $^{12}$CO $J$=2$-$1 model emission.  The spectral averaging 
(middle panel), Hanning smoothing (right panel), and Fourier sampling (bottom 
row) all diminish the apparent morphological asymmetry even though the first 
two effects are operations along the spectral dimension.  The muted asymmetry of
the $^{12}$CO $J$=2$-$1 emission can be attributed to the larger beam size and 
coarser spectral resolution of the Band 6 observations and {\it not} to the disk
structure (as is the case for the rarer $^{13}$CO and C$^{18}$O isotopologues). 
The Band 6 data have been averaged and smoothed in the spectral dimension (see 
\S\ref{sec:models}), which smooths the correlated emission in the spatial 
dimension \citep{thompson86}.  The consequence of a coarse spectral resolution 
is worth emphasizing, since in practice it may significantly degrade the utility
of the excellent spatial resolution of ALMA.

We explored the sensitivity of the Band 7 observations to the gas kinematics
with a series of models where only the bulk velocities were changed. 
We found that these data could distinguish between models for which those gas 
velocity differences were $\gtrsim 5$\%, corresponding to classes of models that
incorporated the vertical geometry of the disk as well as the radial pressure 
gradient in the gas.  Indeed, a few disks have been identified as 
having sub-Keplerian velocities \citep{wang12}, and we suggest that these 
phenomena provide an intuitive explanation (although might not be sufficient for
such systems).  For our fiducial model, the disk self-gravity did not appear to 
significantly alter the velocity field, but may be important when modeling the 
density structure.  Furthermore, the contributions from the gas pressure 
gradient and self gravity are model dependent, and so may become even more 
important for certain systems \citep[e.g.,][]{cesaroni05,bergin13}.

There are a few aspects of the disk hosted by HD 163296 that accentuate the 
subtle effects we have addressed.  First, the disk is both large and relatively 
nearby so that the longest baselines in this SV dataset ($\sim 400$k$\lambda$) 
correspond to spatial scales that will only be attained by full ALMA for 
T Tauri disks in more distant star-forming clusters.  Second, HD 163296 
is an A star that can passively heat its disk out to large radii.  And so in 
addition to having a large radial extent and thick vertical geometry, the disk 
is warm and bright for sub-millimeter observations.  Lastly, the intermediate 
disk inclination helps to evenly project both the vertical and radial dimension 
of the disk on the sky plane \citep{semenov08}.  The inclination also projects 
a significant fraction of the intrinsic gas motions along the line of sight.
This reduces the smoothing effect that the spectral resolution has upon the 
spatial pattern of the line (see Figure \ref{fig:spec_effects}). 

With high-resolution, sensitive instruments such as ALMA and the JVLA, there is 
a host of opportunities for studying circumstellar disks and phenomena 
associated with planet formation
\citep[e.g.][]{wolf05,semenov08,cossins10,cleeves11,gonzalez12,ruge13}.  These 
observations demonstrate the necessity of more sophisticated, physically 
self-consistent approaches when analyzing data from this new generation of 
observing facilities.

\section{Summary}\label{sec:summary}
We have analyzed sensitive, sub-arcsecond resolution observations of
the $^{12}$CO $J$=3$-$2, $^{12}$CO $J$=2$-$1, $^{13}$CO $J$=2$-$1, and 
C$^{18}$O $J$=2$-$1 emission lines from the protoplanetary disk hosted by HD 
163296.  The key conclusions of our analysis are:
\begin{enumerate}
\item The $^{12}$CO $J$=3$-$2 spectral line features a clear and systematic
morphological and brightness asymmetry across the major axis of the disk.  The 
$^{12}$CO $J$=2$-$1 line exhibits similar, but muted, behavior.  No asymmetries
are observed for the rarer $^{13}$CO and C$^{18}$O isotopologues.
\item  The resolved morphological and brightness asymmetries of the 
$^{12}$CO $J$=3$-$2 line emission as well as the symmetric emission of the 
rarer isotopologues are the signatures of a vertical temperature gradient in the
disk.  A double cone structure encapsulates the salient physical features of a 
disk with a cold midplane and warm atmosphere and mimics the distinctive
emission morphology by moving the surface that dominates the emission above 
(and below) the disk midplane.
\item We presented a series of simple disk structures that demonstrated how the
observed data can be roughly reproduced by a disk with a cold midplane and 
warm atmosphere.  For our fiducial model, the $\tau=1$ surface for the 
$^{12}$CO emission was suspended $\sim 15^\circ$ above the midplane, agreeing 
with our morphological characterization using the toy double cone structure.
\item The excellent spatial and spectral resolution of the Band 7 observations
are suprisingly sensitive to the bulk velocities of the gas and can distinguish 
between models that include the vertical geometry of the disk
and radial pressure gradient (a fractional difference in the bulk gas velocity 
field of $\gtrsim 5$\%).  The inclusion of self-gravity for our fiducial disk 
model is a less important correction, but may become dominant for other disk 
models.
\item  The excellent sensitivity and spatial resolution of these ALMA SV data
require careful processing of models.  In particular, coarse spectral 
resolution of the data can strongly impact the spatial morphology of the 
observed emission.
\end{enumerate}

\acknowledgments 
We are grateful to Matt Payne, Diego Mu{\~n}oz, Eugene Chiang, David Knezevic, 
Moritz G{\"u}nther, Marc Metchnik, and Kees Dullemond for insightful 
conversations.  This paper makes use of the following ALMA data: 
ADS/JAO.ALMA\#2011.0.00010.SV.  ALMA is a partnership 
of ESO (representing its member states), NSF (USA) and NINS (Japan), together 
with NRC (Canada) and NSC and ASIAA (Taiwan), in cooperation with the Republic 
of Chile.   The Joint ALMA Observatory is operated by ESO, AUI/NRAO and NAOJ.  
We acknowledge support from NASA Origins of Solar Systems grant No. NNX11AK63.

\clearpage

\begin{deluxetable}{lcccc}
\tablecolumns{5}
\tablewidth{0pc}
\tablecaption{Emission Line Results \label{tab:lines}}
\tablehead{
Parameters & $^{12}$CO(3-2) & $^{12}$CO(2-1) & $^{13}$CO(2-1) & C$^{18}$O(2-1)
}
\startdata
Channel Width\tablenotemark{a} [km s$^{-1}$] & 0.11 & 0.32 & 0.33 & 0.33 \\
Beam & $0\farcs65\times0\farcs42$ & $0\farcs81\times0\farcs66$ & $0\farcs87\times0\farcs70$ & $0\farcs87\times0\farcs70$ \\
P.A. & $-87\fdg3$ & $75\fdg8$ & $76\fdg8$ & $76\fdg8$\\
RMS [Jy beam$^{-1}$] & 0.022 & 0.016 & 0.009 & 0.008 \\
Integrated intensity [Jy km s$^{-1}$] & $109\pm11$ & $46\pm5$ & $18\pm2$ & $5.8\pm 0.6$\\
Peak flux [Jy beam$^{-1}$]& 0.86 & 0.87 & 0.45 & 0.23 \\
\enddata
\tablenotetext{a}{The velocity resolution indicated by the channel width 
is smaller than the true velocity resolution as the data are Hanning 
smoothed along the spectral dimension \citep{lundgren12}.}
\end{deluxetable}

\end{document}